\documentclass[prd,twocolumn,showpacs,showkeys,preprintnumbers,floatfix,
nofootinbib, superscriptaddress]{revtex4-1}
\usepackage{amsfonts} 
\usepackage{amssymb} 
\usepackage{amsmath} 
\usepackage{graphicx} 
\usepackage{subfigure} 
\usepackage{array} 
\usepackage{dcolumn} 
\usepackage{bm} 
\usepackage{latexsym} 
\usepackage{longtable} 
\usepackage{hyperref} 
\graphicspath{{./figs/}}
\renewcommand{\tilde}{\widetilde}
\begin{document}


\title{One-loop matching of improved four-fermion staggered operators
with an improved gluon action}
\author{Jongjeong Kim}
\email[Email: ]{rvanguard@gmail.com}
\affiliation{
  Physics Department,
  University of Arizona,
  Tucson, AZ 85721, USA
}
\author{Weonjong Lee}
\email[Email: ]{wlee@snu.ac.kr}
\homepage[Home page: ]{http://lgt.snu.ac.kr/}
\altaffiliation[Visiting professor at ]{
 Physics Department,
 University of Washington,
 Seattle, WA 98195-1560, USA
}

\affiliation{
 Lattice Gauge Theory Research Center, FPRD, and CTP \\
 Department of Physics and Astronomy,
 Seoul National University, 
 Seoul, 151-747, South Korea
}
\author{Stephen R. Sharpe}
\email[Email: ]{sharpe@phys.washington.edu}
\homepage[Home page: ]{http://www.phys.washington.edu/users/sharpe/}
\affiliation{
 Physics Department, University of Washington, 
 Seattle, WA 98195-1560, USA \\
}
\collaboration{SWME Collaboration}
\date{\today}
\begin{abstract}

 We present results for one-loop matching factors of four-fermion
 operators composed of HYP-smeared staggered fermions.
 We generalize previous calculations by using the tree-level
 improved Symanzik gauge action. These results are needed for our
 companion numerical calculation of $B_K$ and related matrix elements.
 We find that the impact on one-loop matching factors
 of using the improved gluon action is much smaller than that
 from the use of either HYP smearing or mean-field improvement.
 The one-loop coefficients for mean-field improved, 
  HYP-smeared operators
 with the Symanzik gauge action have a maximum magnitude of 
 $O(1)\times \alpha_s$, indicating that perturbation theory
 is reasonably convergent.
\end{abstract}
\maketitle

\section{Introduction \label{sec:intr}}

Numerical simulations of lattice QCD are now able to
calculate a range of phenomenologically interesting non-perturbative
quantities with high precision.
Of particular interest are
hadronic matrix elements of operators that appear in the 
electroweak Hamiltonian, or in extensions of the standard model.
For such quantities it is necessary (in order to make use of
Wilson coefficients calculated in continuum perturbation theory)
to determine the matching factors which relate operators regularized
on the lattice with those regularized in the continuum. 
For the latter one typically uses naive dimensional regularization (NDR) 
with $\overline{\rm MS}$ subtraction. 

In this paper we calculate matching factors (which are, in general,
matrices)
for four-fermion operators composed of light staggered quarks.
These arise, for example,
in the calculation of the $K^0-\overline{K}^0$ mixing 
parameter $B_K$. Their flavor structure forbids mixing
with lower-dimensional operators, so the matching is only between
operators of dimension 6. In the electroweak theory, the operator
that arises has a ``left-left'' structure, due to the left-handed
coupling of the W-bosons. In extensions of the standard model, however,
$\Delta S=2$ operators can arise with other Dirac structures. For this
reason we calculate matching factors for all possible Dirac structures.

In recent years, it has become increasingly common to determine
matching factors non-perturbatively, either using the Rome-Southampton
non-perturbative renormalization method~\cite{Martinelli:1994ty}, 
or using approaches based on the Schr\"odinger 
functional~\cite{Jansen:1995ck}. 
These methods replace
hard-to-estimate truncation errors by controllable statistical and
systematic errors. We are implementing such calculations for
improved staggered fermions, but have so far only
obtained results for bilinear operators~\cite{Lytle:2009xm}.
We expect that the implementation for four-fermion operators,
which involves mixing with a long list of lattice operators, will
be more challenging. The use of one-loop matching is a useful
intermediate step, and, as we will describe, the necessary
calculations are relatively simple generalizations of previous work.
We also note that, since our present numerical calculations involve
very small lattices spacings ($a\approx 0.045\;$fm), the truncation
errors are quite small, since they
are proportional to $\alpha_s^2$ with $\alpha_s$
evaluated at a scale $\approx 1/a$~\cite{Yoon:2010bm}.

Our companion numerical calculations use valence staggered fermions which
have been improved by the use of HYP-smeared links
(links replaced with hypercubic blocked links~\cite{Hasenfratz:2001hp}).
The ensembles are those generated by the 
MILC collaboration~\cite{Bernard:2001av},
in which the gauge action is Symanzik-improved,
and the quark action is the asqtad staggered action.
Previous calculations have obtained the matching factors
for four-fermion operators composed
of HYP-smeared staggered fermions~\cite{Lee:2003sk}, 
but only using the Wilson gauge action.
Here we generalize these results to the case of an improved
gauge action. This extends our earlier work in which
we calculated matching factors
for bilinear operators using improved gauge actions and HYP-smeared staggered
fermions \cite{Kim:2010fj}.

At first sight, the generalization from the Wilson gauge action
(for which, in Feynman gauge, the gluon propagator is diagonal in Euclidean
indices) to an improved gauge action (in which the propagator is
not diagonal) appears non-trivial. In particular, one-loop
calculations using unimproved staggered fermions
\cite{Patel:1992vu,Sharpe:1993ur}
were simplified using the diagonal nature of the gluon propagator.
The inclusion of smeared links, however, leads to the
natural introduction of a ``composite gluon propagator'' which
represents both the smearing and the gluon propagator itself.
This propagator contains non-zero off-diagonal elements, and so
calculations of one-loop matching factors for HYP-smeared
staggered fermions must already deal with the presence
of such elements~\cite{Lee:2002ui,Lee:2003sk}. 
This means that the generalization to an improved gluon
propagator requires no change to the analytic expressions---all
one needs to change is the composite gluon propagator
before numerical evaluation of the loop integral.
As we discuss here, this simplification holds not only
for bilinear operators~\cite{Kim:2010fj}, but also for
four-fermion operators.

The paper is organized as follows.
In Sec.~\ref{sec:act-op}, we recall our notation and 
conventions for actions and operators.
In Sec.~\ref{sec:Feyn}, we present the Feynman diagrams
and describe their evaluation.
Because we are building on the work of Refs.~\cite{Lee:2003sk}
and~\cite{Kim:2010fj}, we provide only a minimal discussion of 
technical details.
In Sec.~\ref{sec:matching}, we present our numerical results 
for matching factors, providing the complete matching matrix
for the operators relevant to $B_K$, and a partial matrix
(the part that will likely be used in practice) for other
four-fermion operators.
We conclude briefly in Sec.~\ref{sec:conclude}.

\section{Actions and Operators \label{sec:act-op}}

The HYP-smeared staggered action has the same form as the
unimproved staggered fermion action,
\begin{equation}
 S_\text{HYP} = 
 \sum_{n} \bar{\chi}(n) \Big[
 \sum_{\mu} \eta_{\mu}(n) \nabla_\mu^\text{H} + m \Big]\chi(n) \,,
 \label{eq:SHYP}  
\end{equation}
where $\eta_\mu(n) = (-1)^{n_1 + \cdots + n_{\mu-1}}$,
and the covariant difference operator is 
\begin{equation}
 \nabla_\mu^\text{H} \chi(n) =
 \frac12
 [V_\mu (n) \chi(n + \hat{\mu}) -
 V^{\dagger}_{\mu}(n-\hat{\mu}) \chi(n-\hat{\mu})] \,.
\end{equation}
Here and in the following we set the lattice spacing $a$ to unity, 
except where confusion could arise. 
HYP improvement consists of using HYP-smeared links,
$V_\mu$, instead of the original ``thin'' links, $U_\mu$.
We set the HYP-smearing parameters to the values
that remove the tree-level coupling of quarks to gluons having one or 
more components of momenta equal to $\pi$.
These values are $\alpha_1 = 0.875$, $\alpha_2 = 4/7$ and
$\alpha_3 = 0.25$, in the notation of Ref.~\cite{Hasenfratz:2001hp}.
These are the values used in our numerical simulations.

After gauge-fixing, we expand both the thin and smeared links 
in the usual way,
\begin{eqnarray}
 U_\mu(n) &=& \exp [i g_0 A_\mu (n + \hat{\mu}/2)] \,,
\\
 V_\mu(n) &=& \exp [i g_0 B_\mu (n + \hat{\mu}/2)] \,.
\end{eqnarray}
where $g_0$ is the bare gauge coupling. 
The relation between the fluctuations of smeared and thin links 
can be written
\begin{align}
 B_\mu(n + \hat{\mu}/2) 
 &= \int^\pi_{-\pi}\frac{d^4k}{(2\pi)^4}
 \sum_\nu h_{\mu\nu}(k)
 A_\nu(k)e^{ik\cdot(n + \hat{\mu}/2)} \nonumber\\
 & + \mathcal{O}(A^2) \,.
 \label{eq:b_mn}
\end{align}
Here, $h_{\mu\nu}(k)$ is the smearing kernel, which depends
on the smearing parameters and the details of the HYP
construction.
It contains non-zero off-diagonal components because a smeared
link in one direction contains contributions from thin links in
all four directions.
It turns out that we need only the linear term in Eq.~(\ref{eq:b_mn})
in a one-loop calculation. The contribution of the quadratic
term (which gives rise to tadpole diagrams) turns out to vanish
due to the projection back into the SU(3) group that is part of
the definition of 
HYP-smearing~\cite{Patel:1992vu,Lee:2002fj,Kim:2009tk}. 

The smearing kernel $h_{\mu\nu}$ can be conveniently 
decomposed into diagonal and off-diagonal parts:
\begin{equation}
 h_{\mu\nu}(k) = \delta_{\mu\nu}D_\mu(k)
 + (1 - \delta_{\mu\nu})
 \bar{s}_\mu \bar{s}_\nu
 \tilde{G}_{\nu,\mu}(k) \,,
\label{eq:hmunudef}
\end{equation}
with $\bar{s}_\mu = \sin(k_\mu/2)$, and 
\begin{align}
 D_\mu(k) &= 1 -
 \sum_{\nu\ne\mu} {\bar s}_\nu^2 + \sum_{\nu < \rho \atop
   \nu,\rho\ne\mu}{\bar s}_\nu^2 {\bar s}_\rho^2 - {\bar s}_\nu^2
 {\bar s}_\rho^2 {\bar s}_\sigma^2 \,, \\
 \tilde{G}_{\nu,\mu}(k) & =  1 - \frac{({\bar s}_\rho^2 
     + {\bar s}_\sigma^2)}{2} 
   + \frac{{\bar s}_\rho^2 {\bar s}_\sigma^2}{3} \,. 
\end{align}
Here $\mu$, $\nu$, $\rho$, and $\sigma$ all differ from each other.
By contrast, the smearing kernel for an action
containing the original thin links is simply
$h_{\mu\nu} = \delta_{\mu\nu}$.

We use the tree-level Symanzik-improved gluon action
\cite{Weisz:1982zw,Luscher:1984xn};
\begin{equation}
 S_g = \frac{6}{g_0^2} 
 \bigg[ \frac{5}{3} \sum_{\rm pl} 
\frac{{\rm ReTr} (1 - U_{\rm pl})}{3}
 - \frac{1}{12} \sum_{\rm rt} 
\frac{{\rm ReTr} (1 - U_{\rm rt})}{3}  
 \bigg] \,,
 \label{eq:sg} 
\end{equation} 
where ``pl'' and ``rt'' represent plaquette and rectangle,
respectively. In fact, the MILC collaboration use the (partial)
one-loop Symanzik-improved action determined in
Refs.~\cite{Luscher:1985zq,Alford:1995hw}. However, 
the one-loop contributions to this action contribute 
to matching factors of valence 
fermionic operators only at two-loop level,
so the consistent choice for our one-loop calculation is
the tree-level action (\ref{eq:sg}).
For purposes of comparison, we also use the Wilson gluon action,
which is obtained from Eq.~(\ref{eq:sg}) by dropping
the rectangle term and setting the coefficient of the 
plaquette to unity instead of $5/3$.

Since we use MILC asqtad ensembles in our numerical studies, the sea
quarks are asqtad staggered fermions rather than HYP-smeared.
We do not display the sea-quark action, however, 
since sea-quarks only enter at two-loop order in the matching of 
valence fermionic operators. Our one-loop matching factors are thus
valid for any choice of sea quarks.

We now turn to the definitions of our lattice four-fermion operators,
which are the same as those used in Ref.~\cite{Lee:2003sk}. 
Our construction follows the hypercube method of
Ref.~\cite{KlubergStern:1983dg}. 
The operators come in two classes, differing in the 
contractions of their color indices. First we have
\emph{one color-trace} operators, labeled with a subscript $I$:
\begin{align}
 \lefteqn{[S \times F][S' \times F']_I (y) = } 
 \nonumber\\
 &\frac{1}{4^4}\sum_{A,B,A',B'}
  [\bar{\chi}_a^{(1)}(2y+A)
 \overline{(\gamma_S \otimes \xi_F)}_{AB}
 \chi_b^{(2)}(2y+B)] \nonumber\\
 & \times [\bar{\chi}_{a'}^{(3)}(2y+A')
 \overline{(\gamma_{S'} \otimes \xi_{F'})}_{A'B'}
 \chi_{b'}^{(4)}(2y+B')] \nonumber\\
 & \times 
 \mathcal{V}^{ab'}(2y+A,2y+B')
 \mathcal{V}^{a'b}(2y+A',2y+B) \,.
 \label{eq:1c-tr-op}
\end{align}
Here, $y \in \mathbf{Z}^4$ is the coordinate of $2^4$ hypercubes.  
Hypercube vectors\footnote{%
These are vectors whose entries are $0$ or $1$.}
$S$ and $S'$ denote the spins of the component bilinears,
while $F$ and $F'$ denote the tastes.  Indices $a$, $b$, $a'$, and
$b'$ denote colors, while superscripts $(i)$ for $i=1,2,3,4$ label
different flavors (not tastes). 
Using four different flavors forbids penguin diagrams, which would
lead to mixing with lower-dimension operators.\footnote{%
The relation of these four-flavor operators to the $\Delta S=2$
continuum operators is discussed below.}
Two ``fat'' Wilson lines $\mathcal{V}^{ab'}(2y+A,2y+B')$
and $\mathcal{V}^{a'b}(2y+A',2y+B)$ ensure the gauge invariance of the
four-fermion operators. A fat Wilson line
$\mathcal{V}^{ab'}(2y+A,2y+B')$, for example, is constructed by
averaging over all the shortest paths connecting $2y+A$ and $2y+B'$,
with each path formed by products of HYP-smeared links $V_\mu$. When
we use the unimproved staggered action the Wilson lines are composed
of unsmeared thin links, $U_\mu$.

The second class are the \emph{two color-trace} operators,
for which we use the subscript $II$:
\begin{align}
 \lefteqn{[S \times F][S' \times F']_{II} (y) = }
 \nonumber\\
 & \frac{1}{4^4}\sum_{A,B,A',B'}
 [\bar{\chi}_a^{(1)}(2y+A)
 \overline{(\gamma_S \otimes \xi_F)}_{AB}
 \chi_b^{(2)}(2y+B)] \nonumber\\
 \times 
 & [\bar{\chi}_{a'}^{(3)}(2y+A')
 \overline{(\gamma_{S'} \otimes \xi_{F'})}_{A'B'}
 \chi_{b'}^{(4)}(2y+B')] \nonumber\\
 \times 
 & \mathcal{V}^{ab}(2y+A,2y+B)
 \mathcal{V}^{a'b'}(2y+A',2y+B') \,.
 \label{eq:2c-tr-op}
\end{align}
These operators differ from those with one color-trace 
only by the choice of fat Wilson
lines---here they connect within each bilinear, whereas for the one
color-trace operators they connect between bilinears.

We also consider mean-field improvement of the
staggered action and operators following
Refs.~\cite{Lepage:1992xa,Patel:1992vu,Ishizuka:1993fs,Lee:2001hc}. 
This is also referred to as tadpole improvement. 
Mean-field improvement is achieved by rescaling the staggered
fields and the links. In the case of HYP-smeared staggered fermions
the rescaling is
\begin{align}
 \chi &\to \psi = \sqrt{v_0} \chi \,, \\
 \qquad
 \bar{\chi} &\to \bar{\psi} = \sqrt{v_0} \bar{\chi} \,, \\  
 \qquad
 V_\mu &\to \tilde{V}_\mu = V_\mu / v_0 \,, \\
 \qquad
 v_0 &\equiv \bigg[ \frac13 \text{ReTr} \langle V_{\rm pl}
 \rangle\bigg]^{1/4} \,,
\end{align}
with $V_{\rm pl}$ the plaquette composed of HYP-smeared links.
One then constructs the operators described above out of
$\psi$, $\bar\psi$, and $\tilde{V}_\mu$.
The resulting operators
are expected to be closer to their continuum counterparts
because the rescaled links fluctuate around unity.
Mean-field improvement can be implemented in simulations
after the data has been collected, as long as the contributions
to the four-fermion operators having different numbers of
links are stored separately.

\section{Feynman diagrams and their evaluation\label{sec:Feyn}}

Feynman rules for the gauge and staggered-fermion actions, 
and for insertions of the four-fermion operators, can be found in
literature and we do not reproduce them here. 
The rules for unimproved staggered fermions are given
in Refs.~\cite{Daniel:1987aa,Ishizuka:1993fs,Patel:1992vu},
and the generalization to HYP-smeared staggered fermions can be
found in Ref.~\cite{Lee:2002ui,Lee:2003sk}. 
The gluon propagator for the Symanzik action was determined
in Ref.~\cite{Weisz:1982zw}; we use the simpler form presented in our 
earlier work~\cite{Kim:2009tk}.~\footnote{%
To be precise, 
we use the formulae of Appendix A of Ref.~\cite{Kim:2009tk}
with $\omega=1$, $c=-1/12$ and $c'=0$. The result
for the Wilson gauge action is obtained by further setting
$c=0$.}

\begin{figure}[htbp!]
\subfigure[]{\includegraphics[width=0.23\textwidth]{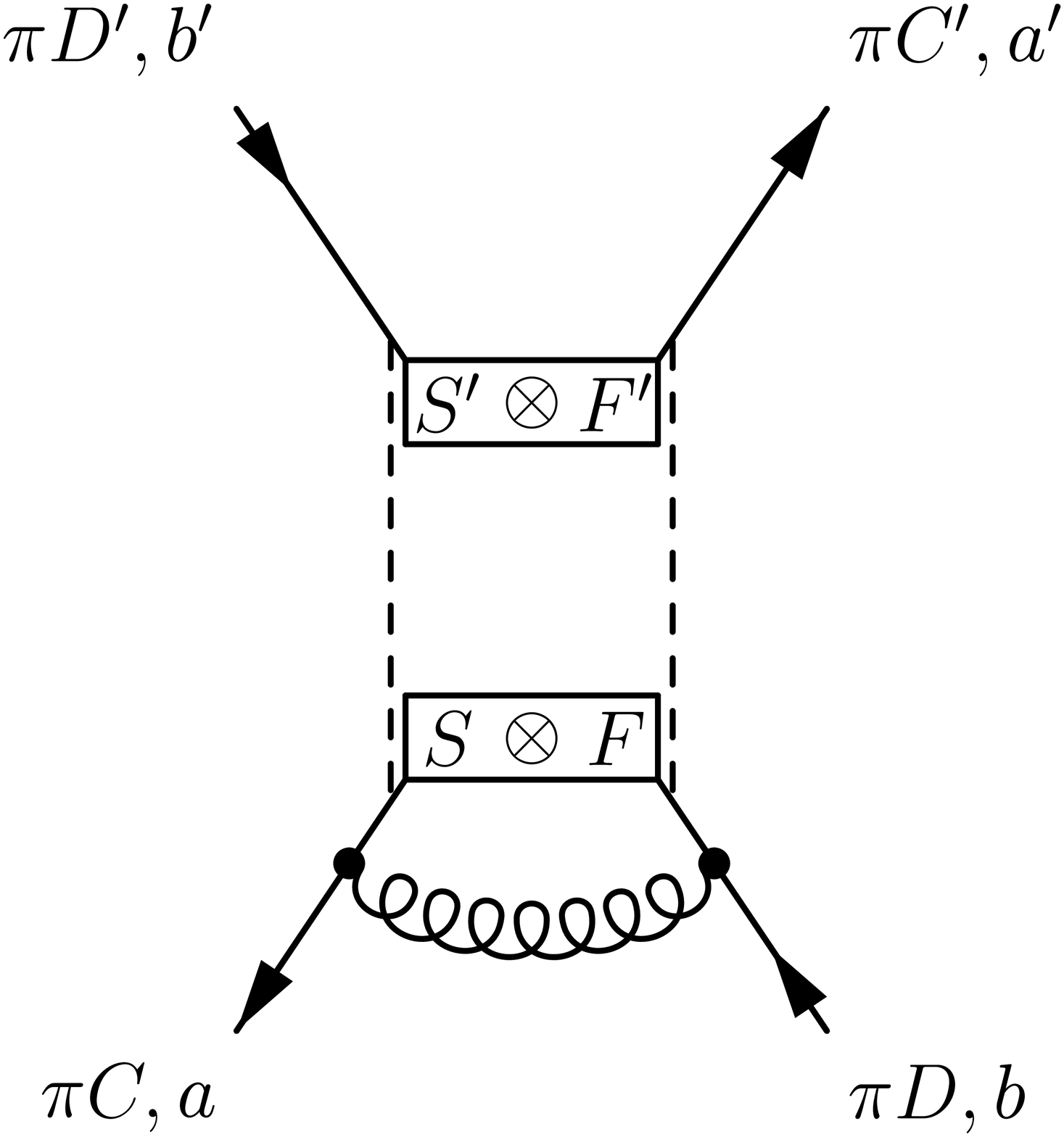}}
\subfigure[]{\includegraphics[width=0.23\textwidth]{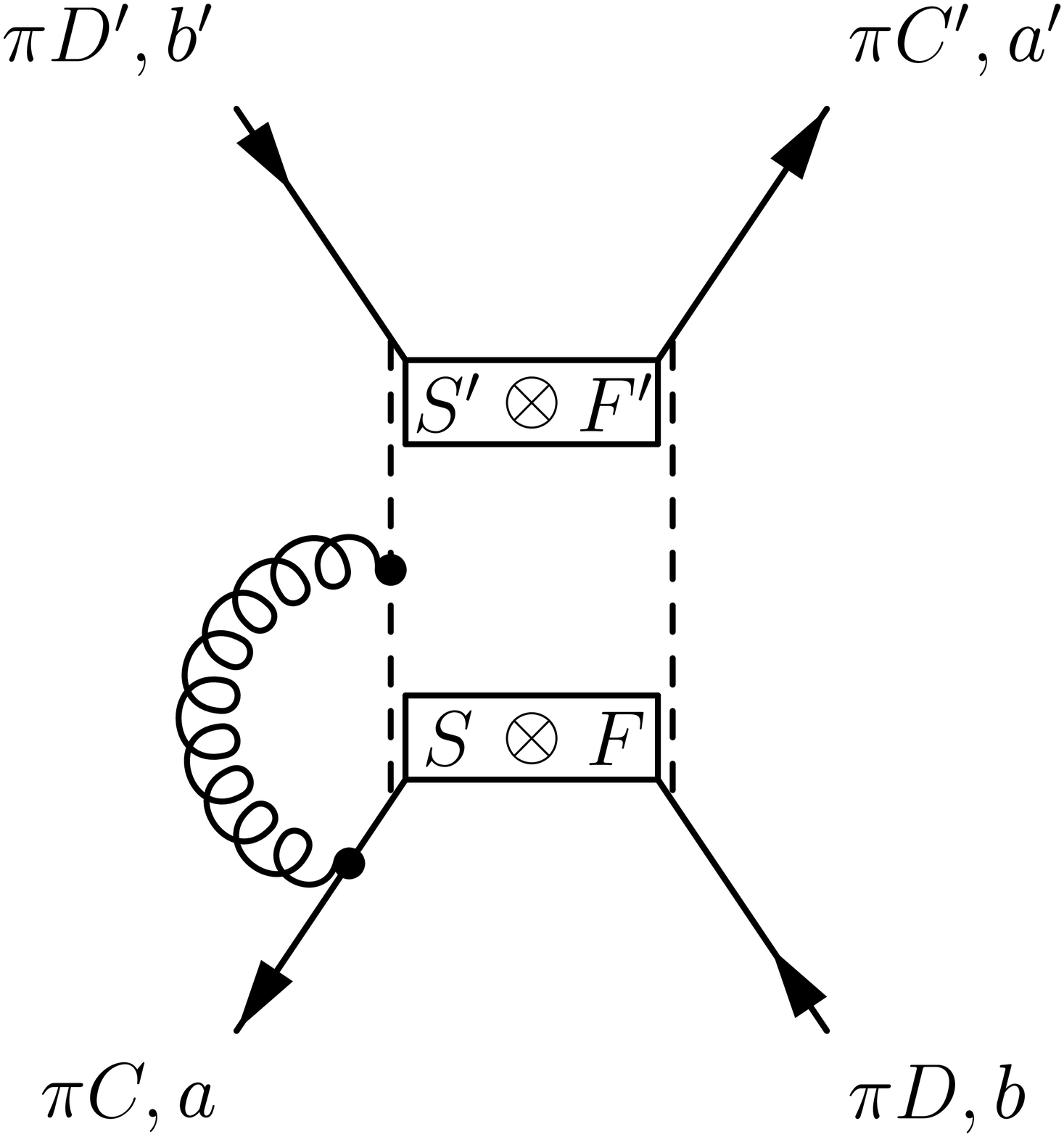}}
\subfigure[]{\includegraphics[width=0.23\textwidth]{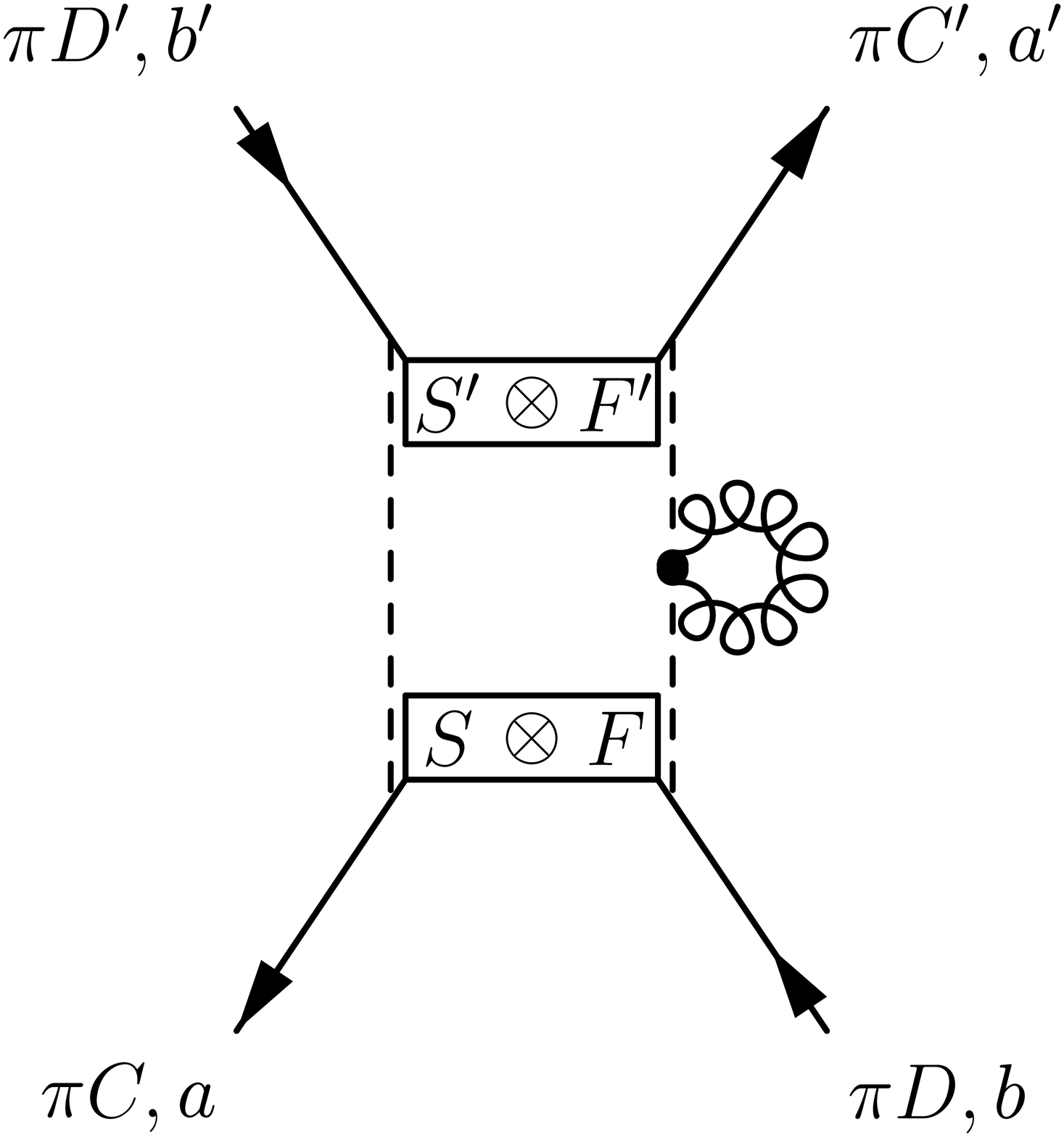}}
\subfigure[]{\includegraphics[width=0.23\textwidth]{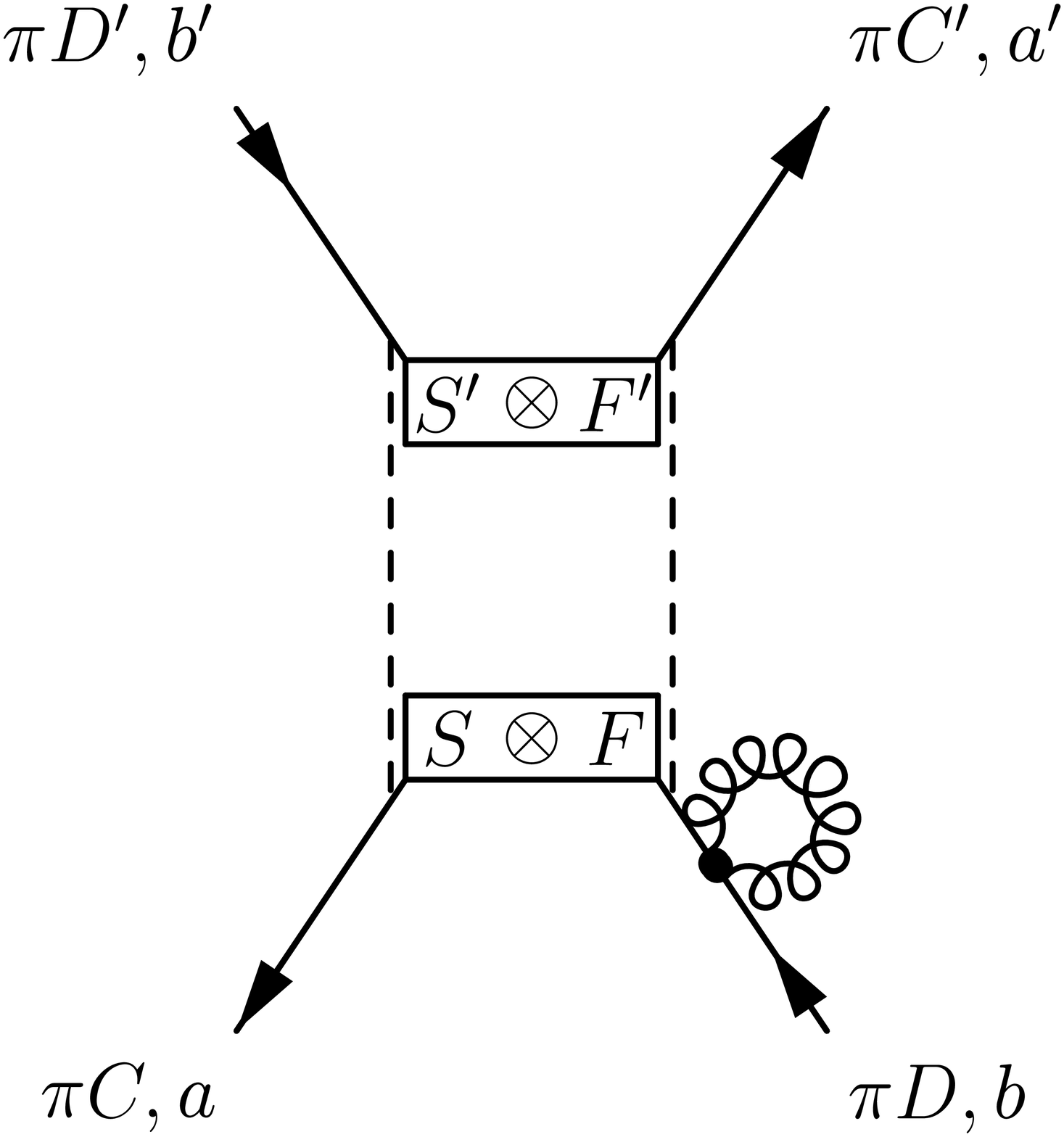}}
\subfigure[]{\includegraphics[width=0.23\textwidth]{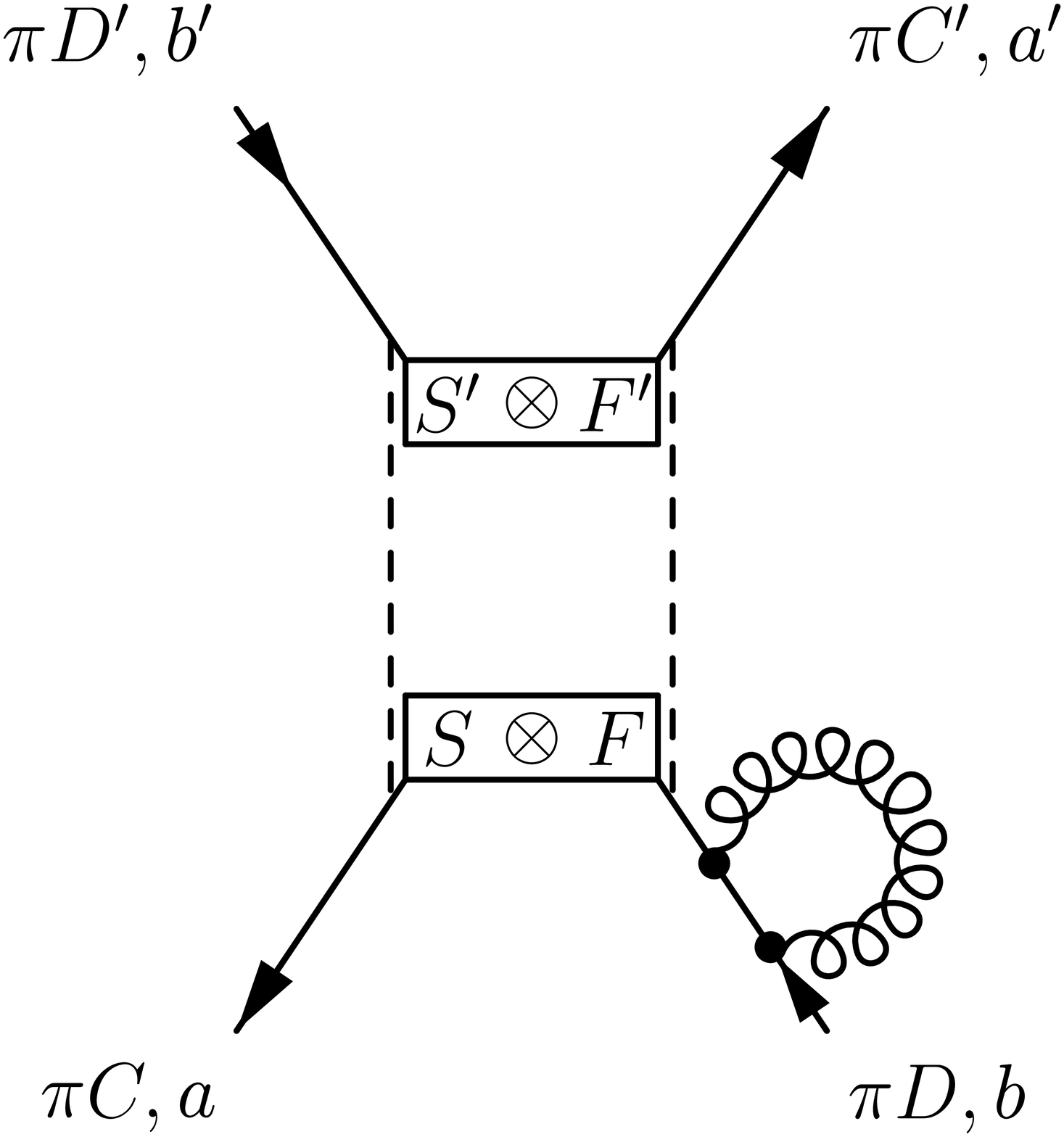}}
\subfigure[]{\includegraphics[width=0.23\textwidth]{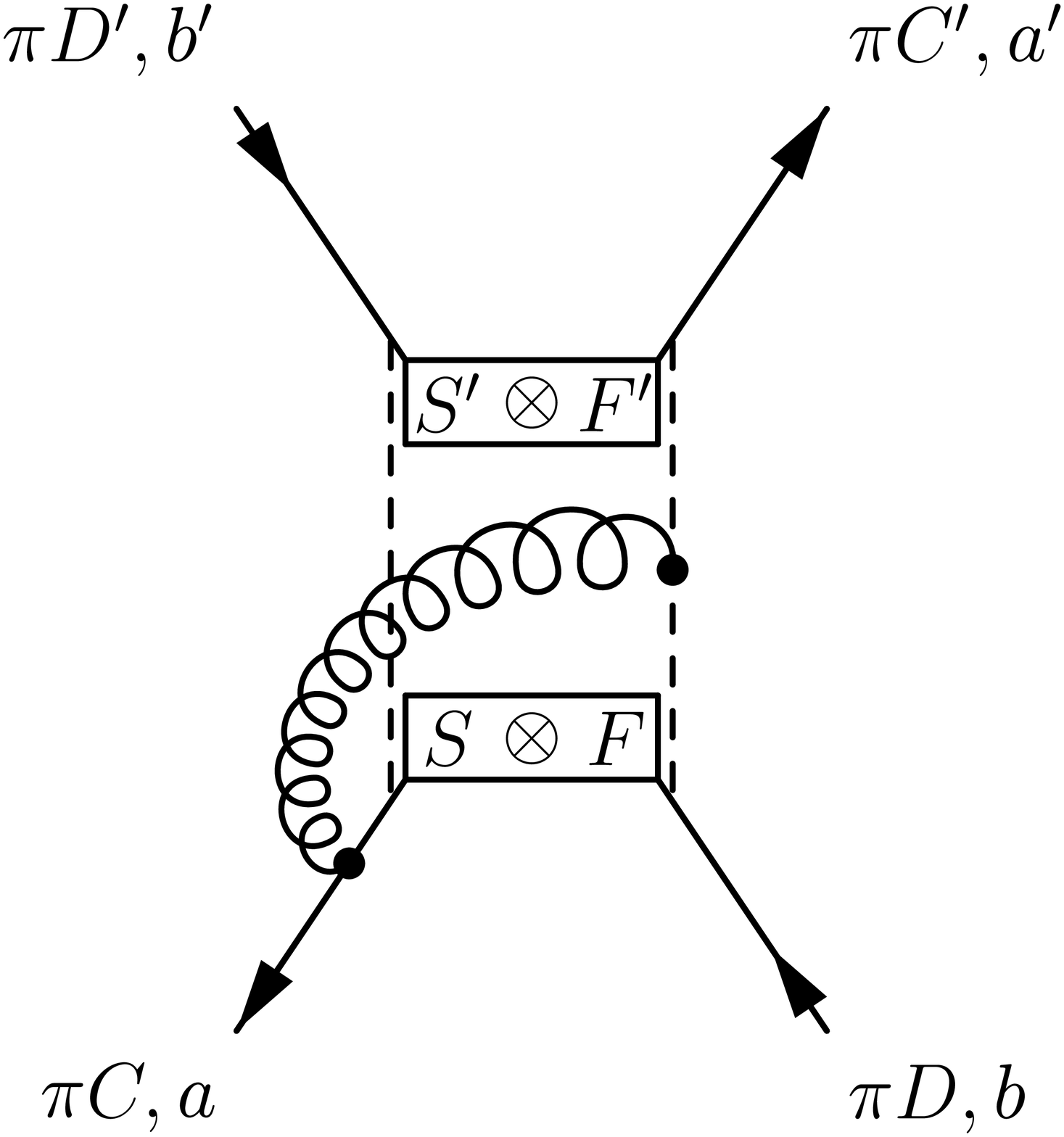}}
\subfigure[]{\includegraphics[width=0.23\textwidth]{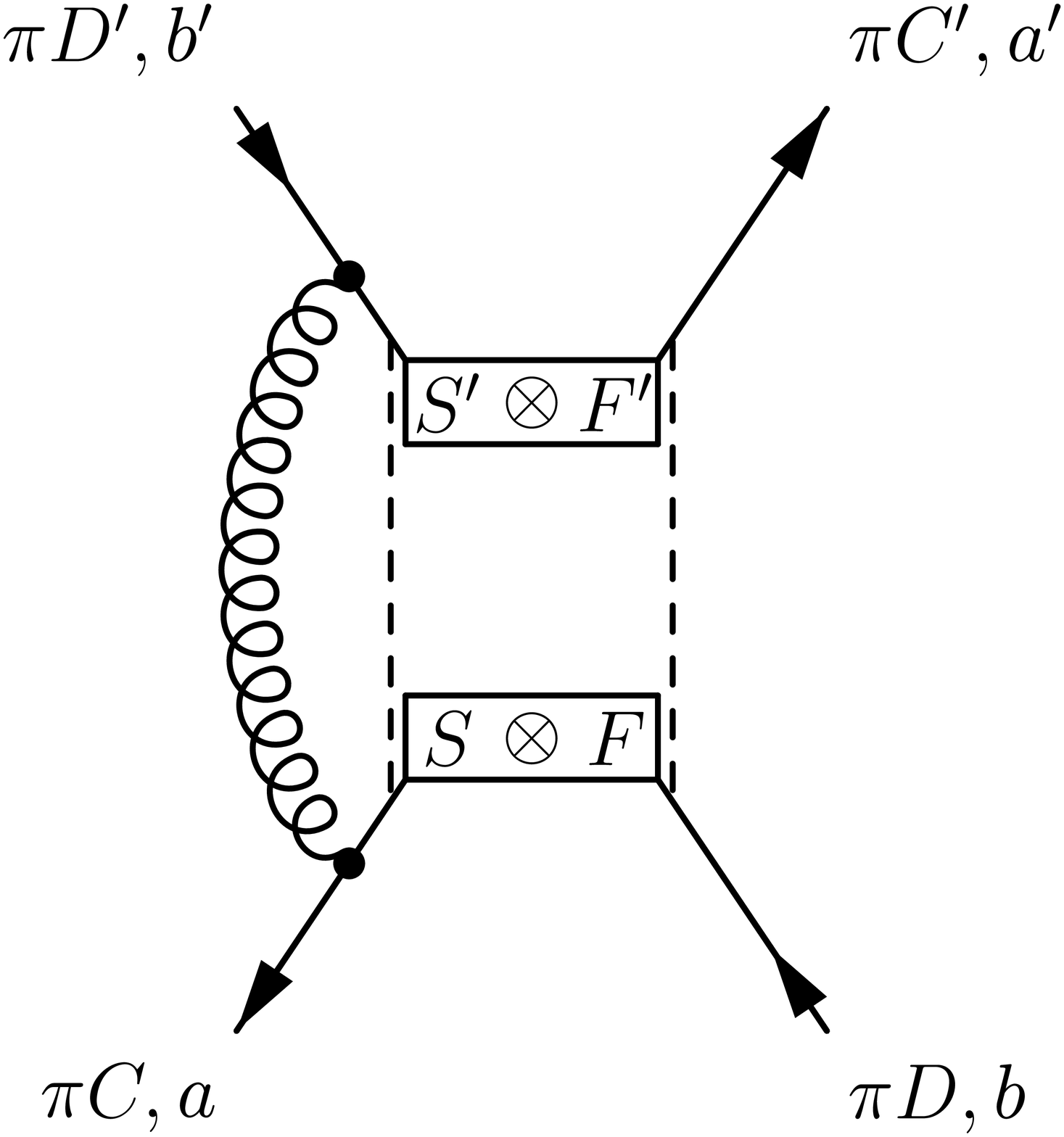}}
\subfigure[]{\includegraphics[width=0.23\textwidth]{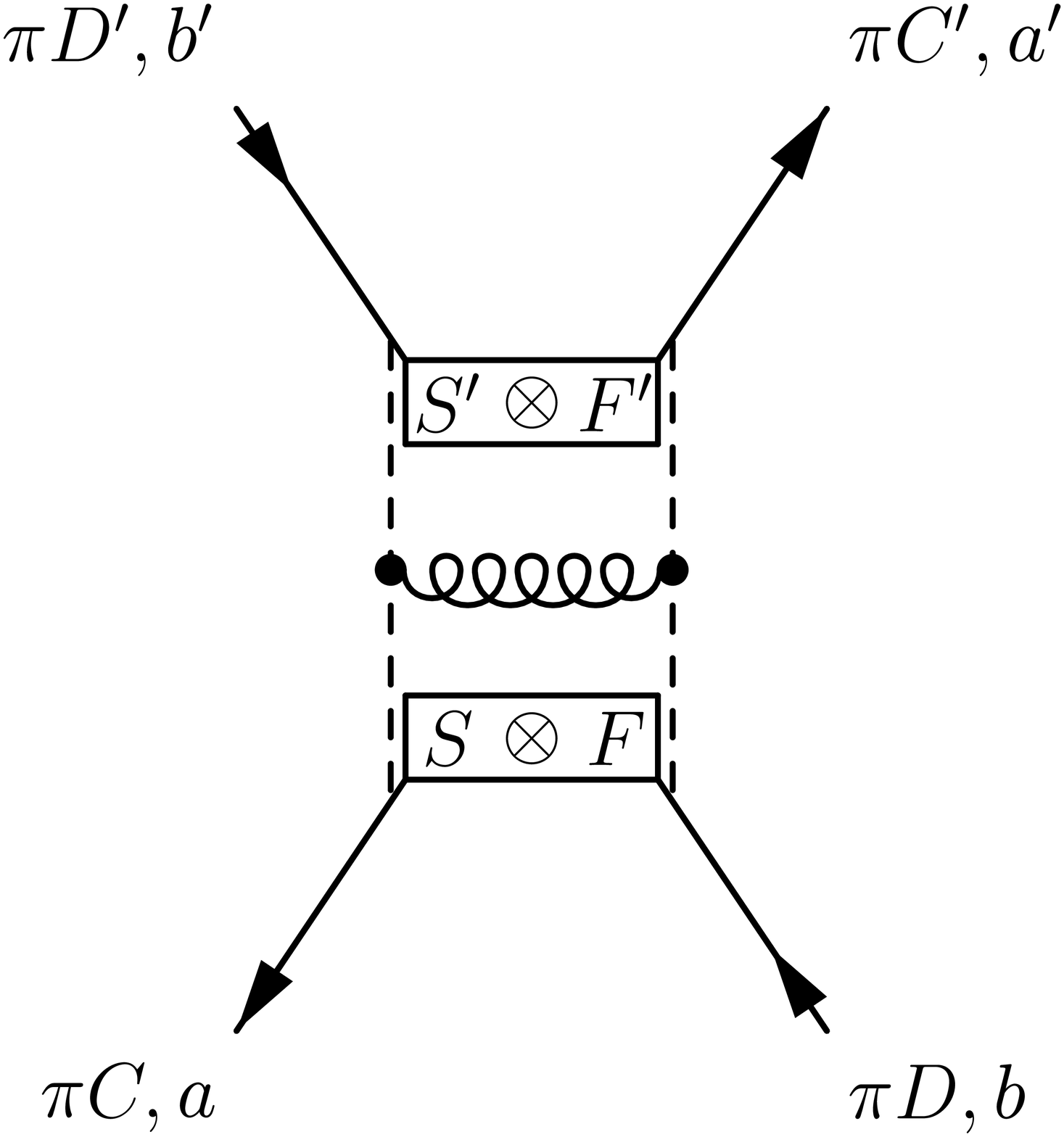}}
\caption{Feynman diagrams contributing to the one-loop
matrix elements of one color-trace operators. 
We show only one diagram of each type. Hypercube vectors ($C$, $D$,
$C'$, and $D'$) multiplied by $\pi$ denote external quark momenta.
$a$, $b$, $a'$, and $b'$ are color indices.
Dashed lines indicate the Wilson lines which make the four-fermion
operator gauge invariant. Boxes indicate the hypercube bilinears
of which the four-fermion operator is composed.
\label{fig:ff1c}}
\end{figure}

\begin{figure}[htbp!]
\subfigure[]{\includegraphics[width=0.23\textwidth]{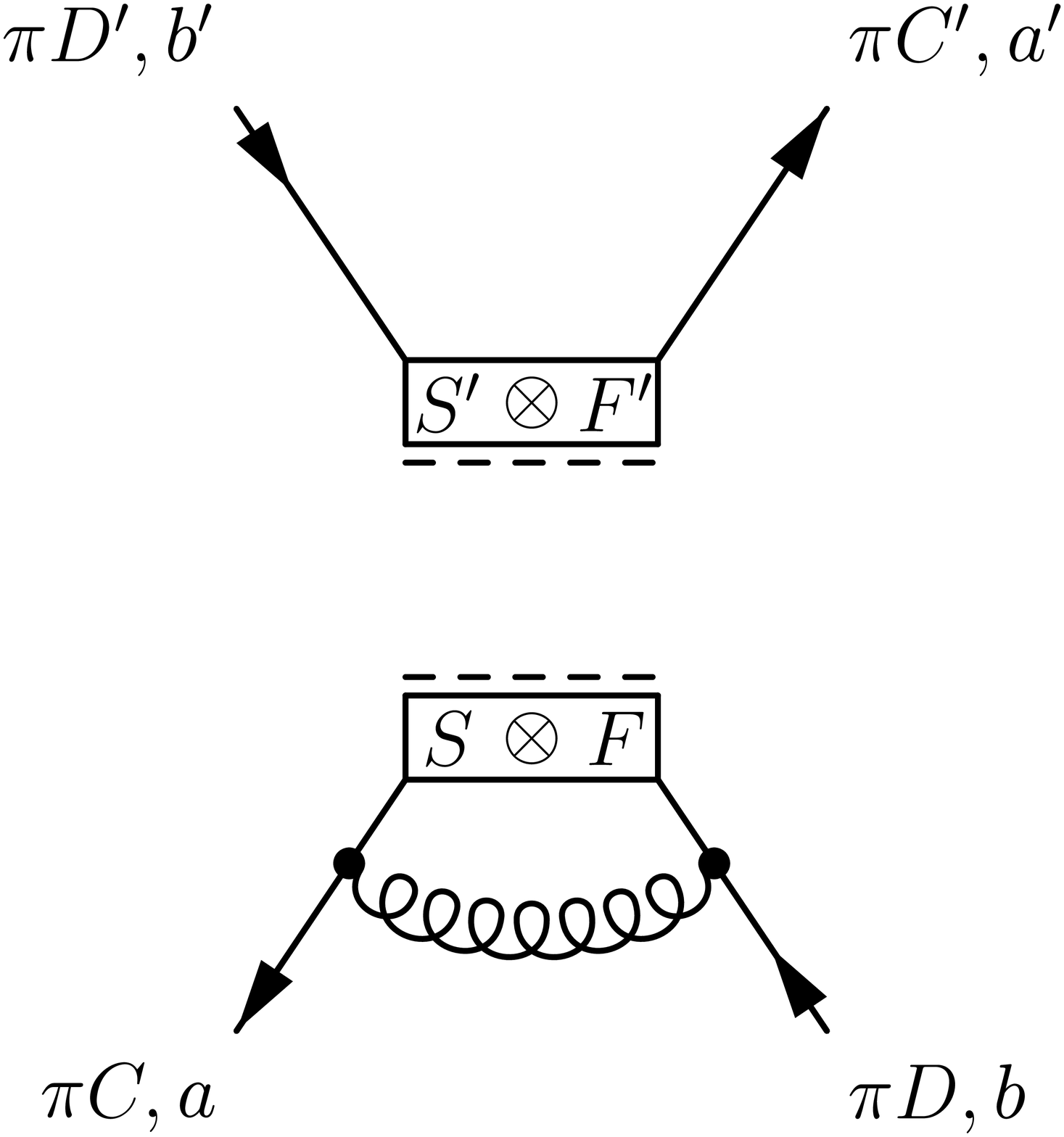}}
\subfigure[]{\includegraphics[width=0.23\textwidth]{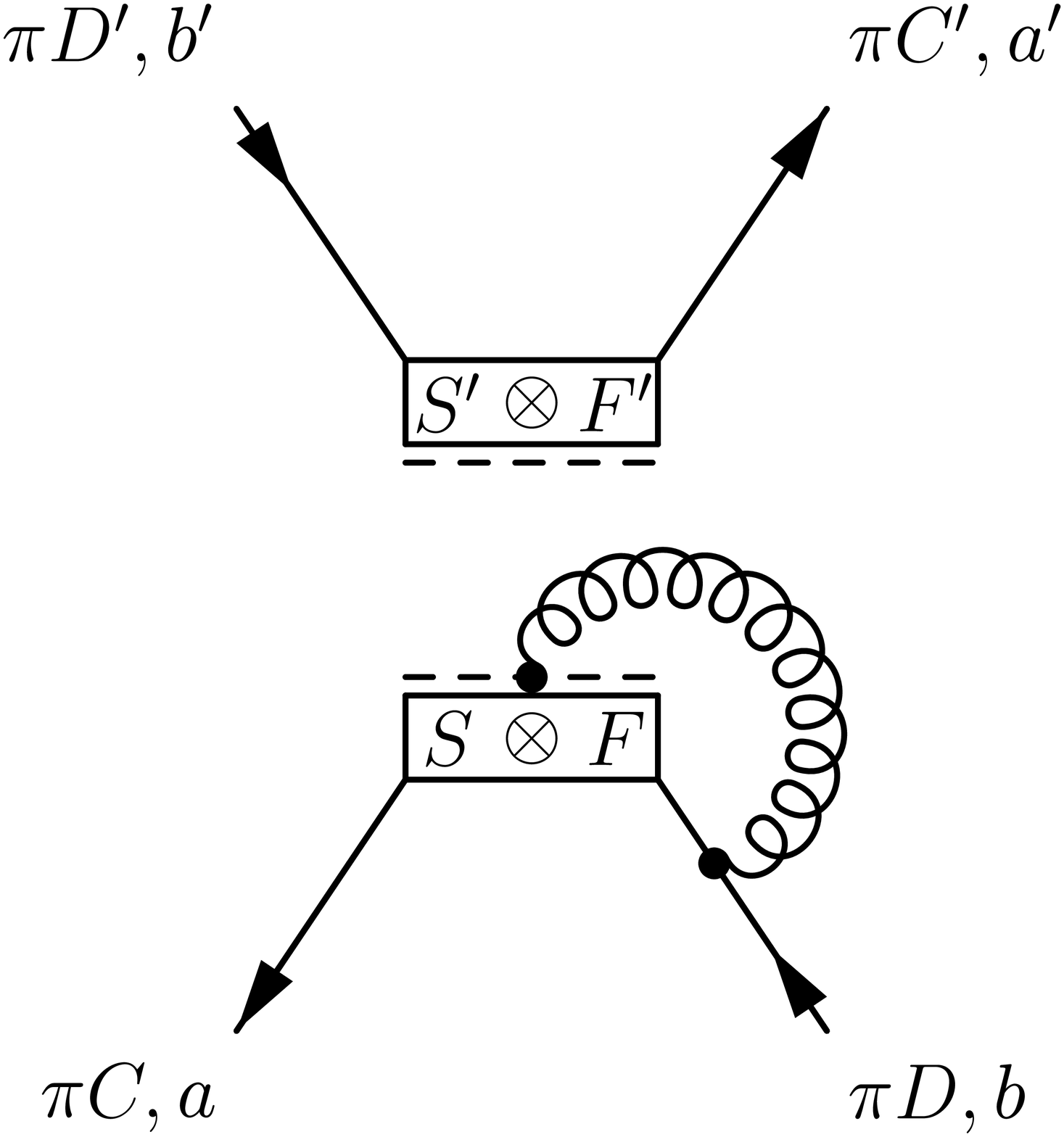}}
\subfigure[]{\includegraphics[width=0.23\textwidth]{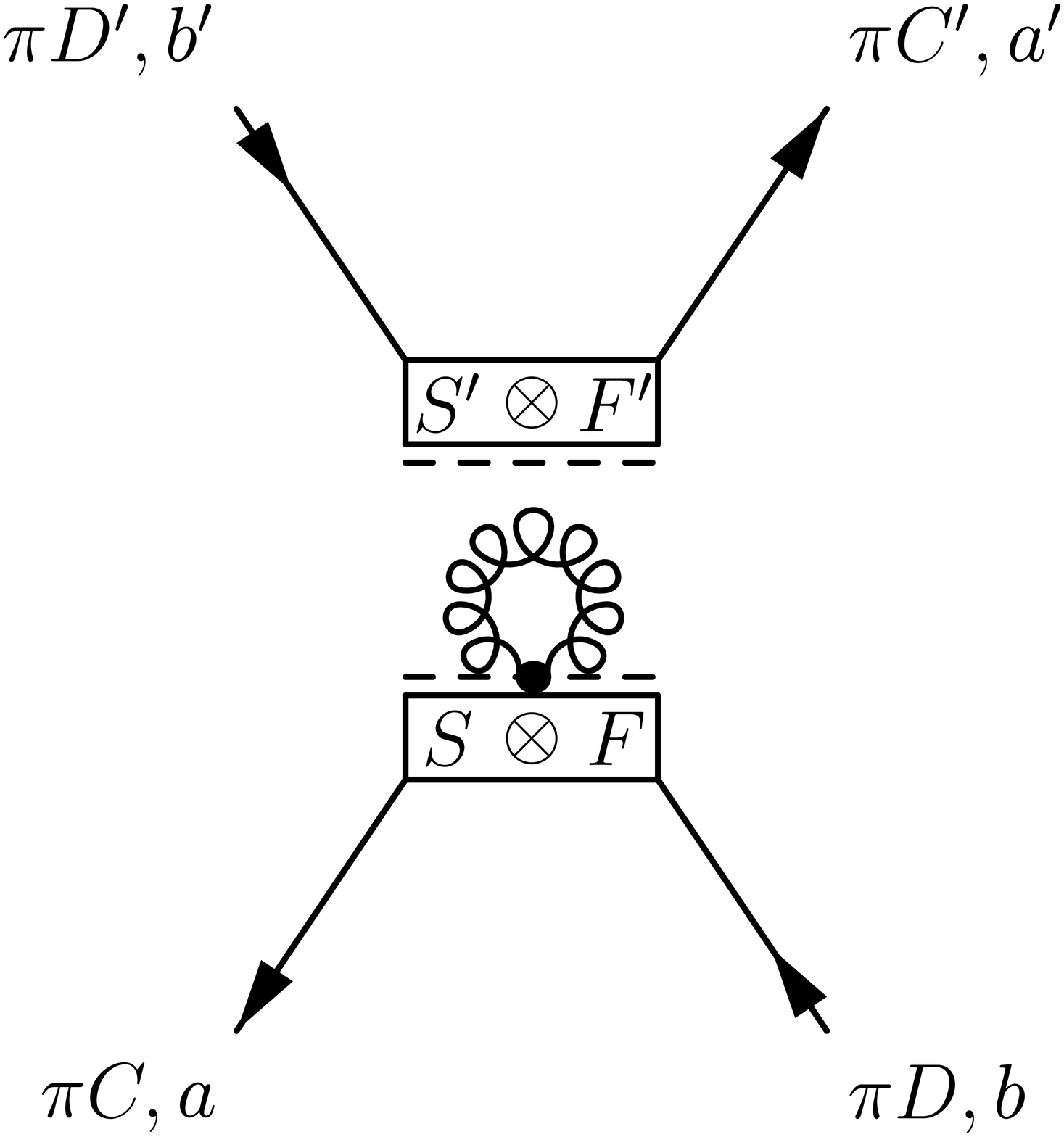}}
\subfigure[]{\includegraphics[width=0.23\textwidth]{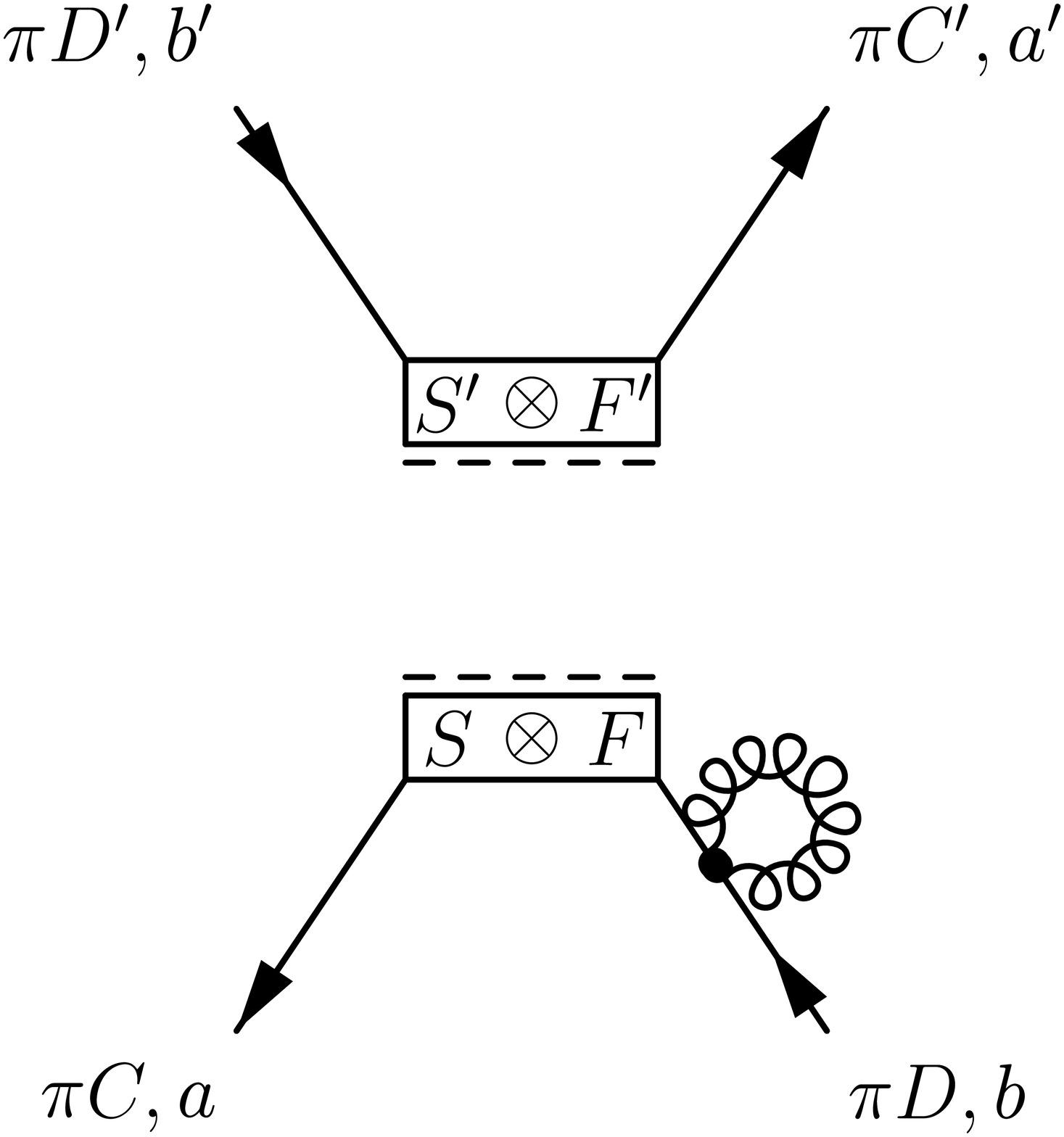}}
\subfigure[]{\includegraphics[width=0.23\textwidth]{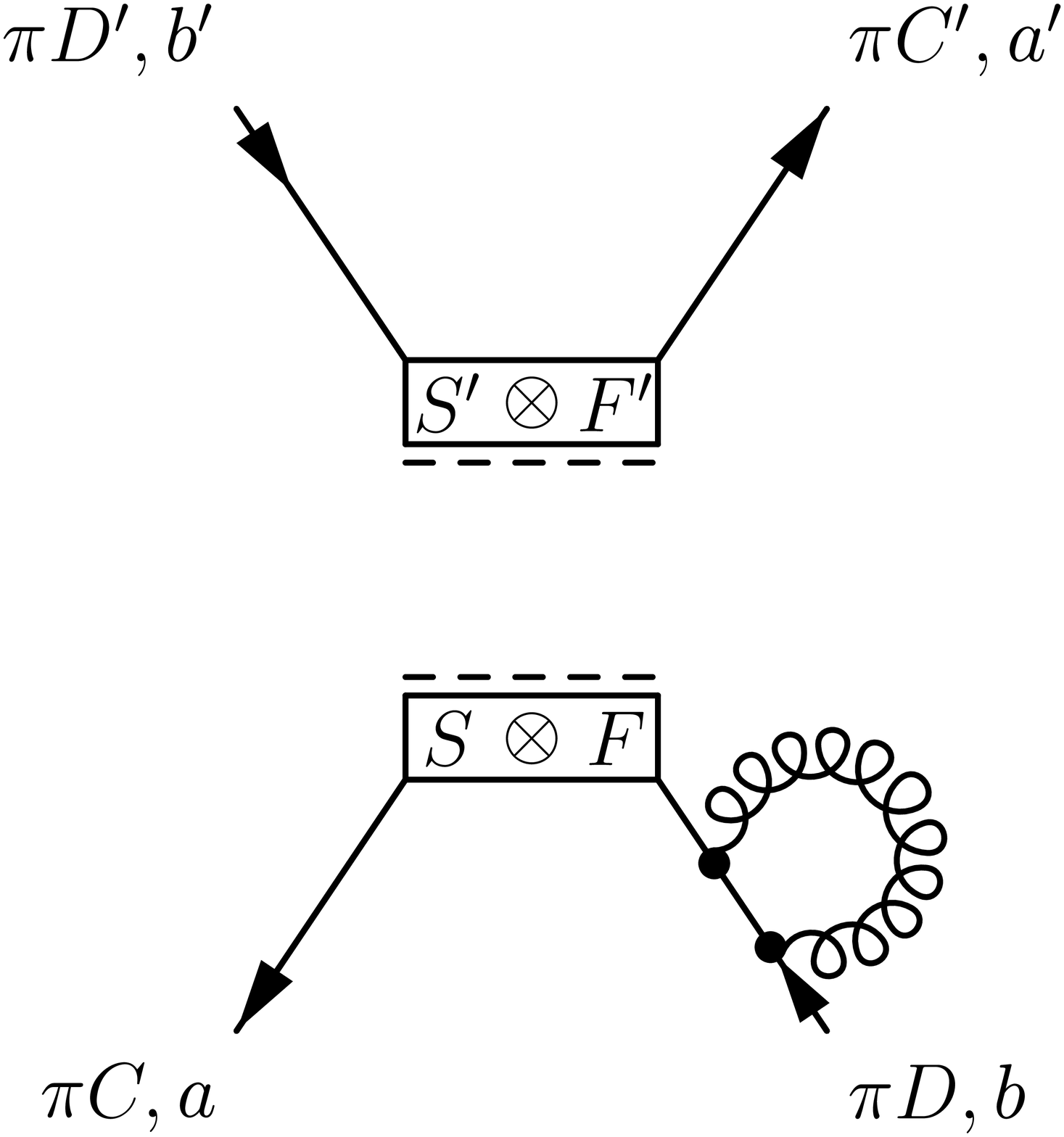}}
\subfigure[]{\includegraphics[width=0.23\textwidth]{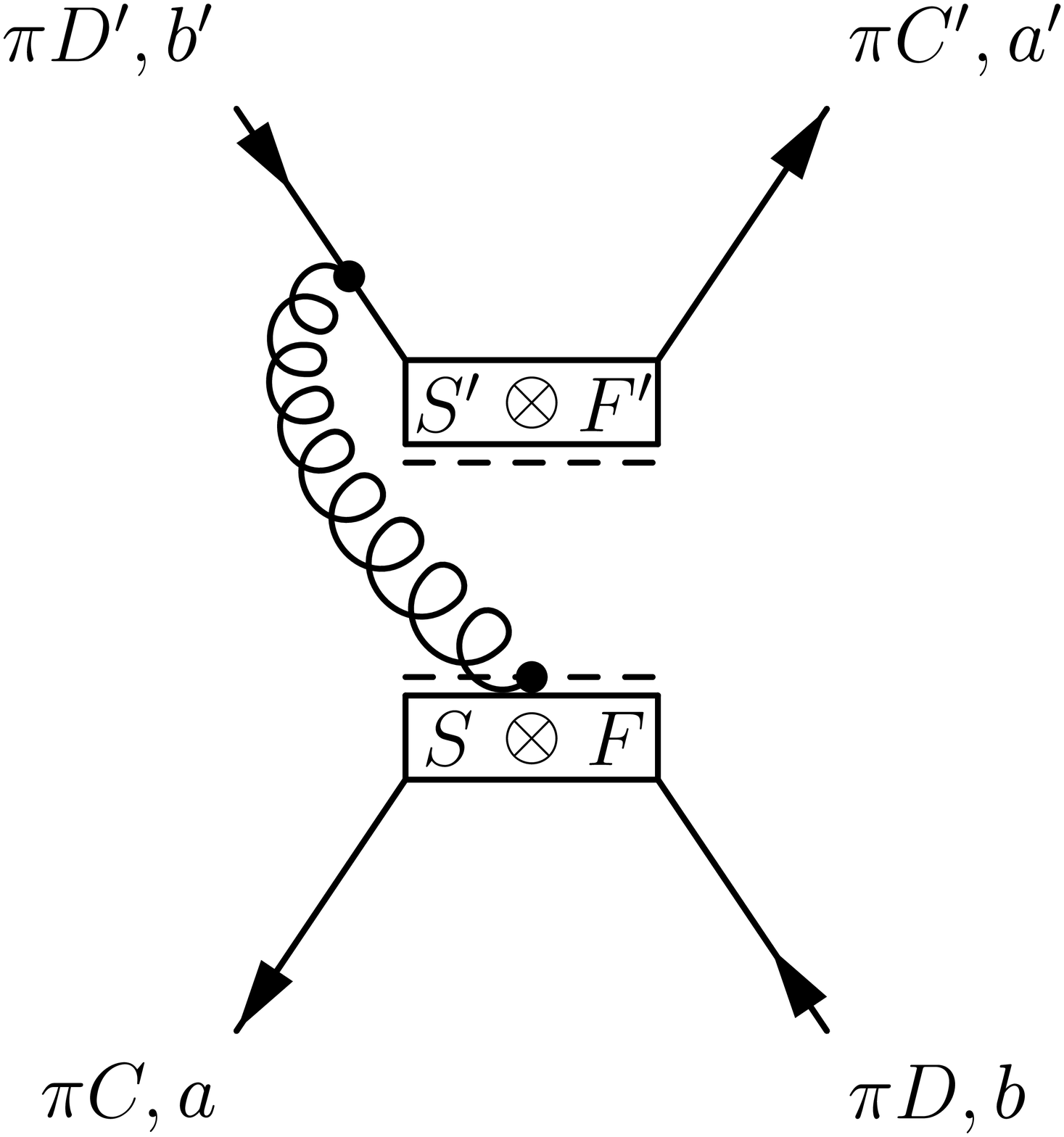}}
\subfigure[]{\includegraphics[width=0.23\textwidth]{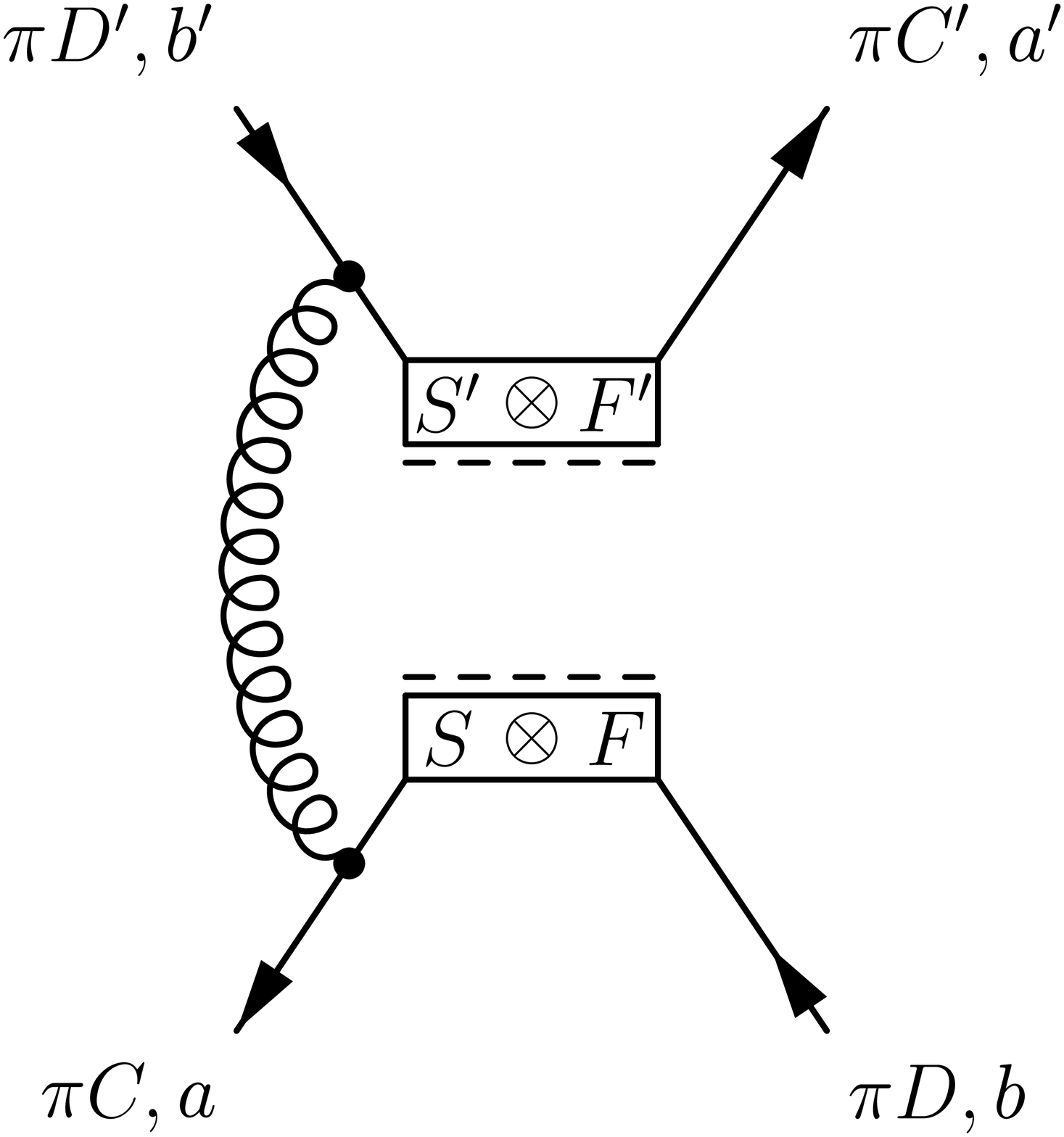}}
\subfigure[]{\includegraphics[width=0.23\textwidth]{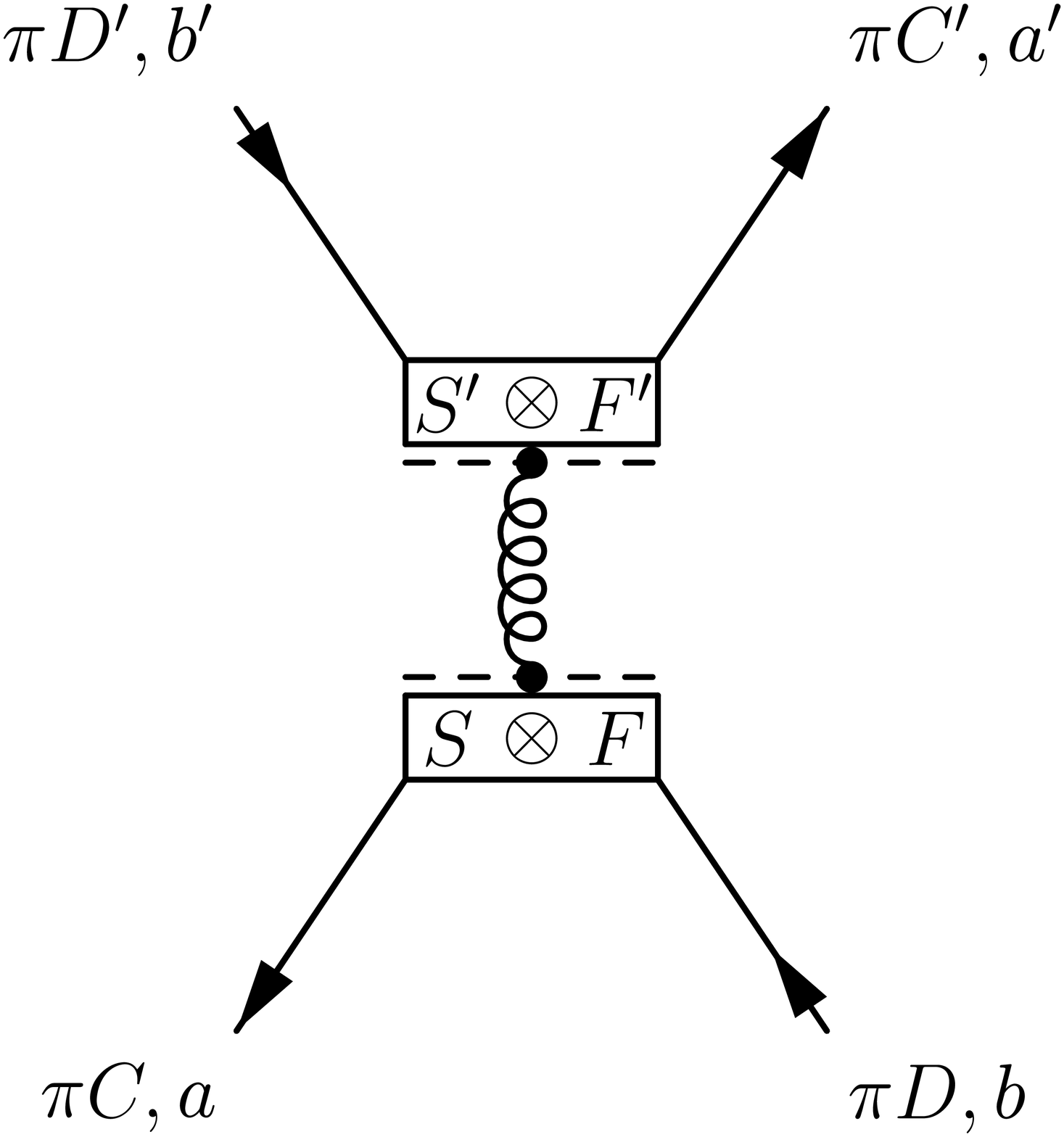}}
\caption{Feynman diagrams contributing to the one-loop
matrix elements of two color-trace operators. 
Notation is as in Fig.~\protect\ref{fig:ff1c}.\label{fig:ff2c}}
\end{figure}

We show the Feynman diagrams contributing to one-loop matching
factors to the two types of four-fermion operators 
in Figs.~\ref{fig:ff1c} and \ref{fig:ff2c}.  
Analytic formulae for these
diagrams for HYP-smeared staggered fermions with the Wilson gluon
action are given in Ref.~\cite{Lee:2003sk}. 
We do not repeat these results here, since it turns
out, as already mentioned in the Introduction, 
that the generalization to the improved gluon action is relatively simple.
Instead we explain the recipe by which the results of Ref.~\cite{Lee:2003sk}
can be generalized.

The key point is that, since all gauge links are
HYP-smeared (whether in the action or the operators), the gluon
propagator always comes with a smearing kernel on each end.
Thus what appears is the \emph{composite gluon propagator} 
(called the ``smeared-smeared propagator'' in Ref.~\cite{Kim:2009tk}):
\begin{equation}
 \mathcal{T}_{\mu\nu}(k) \equiv
 \sum_{\alpha\beta}
 h_{\mu\alpha}(k) h_{\nu\beta}(k)
 \mathcal{D}_{\alpha\beta}(k) \,.
 \label{eq:comp_glu_prop}
\end{equation}
Here $\mu$ and $\nu$ are the directions of the initial and final
smeared gauge links, 
$h$ is given in Eq.~(\ref{eq:hmunudef}),
and $\mathcal{D}_{\alpha\beta}(k)$ is the gluon propagator in Feynman gauge.
Even with the Wilson gauge action, where $\mathcal{D}$ is diagonal,
the fact that $h$ has non-vanishing off-diagonal elements implies that
$\mathcal{T}$ does too.
Thus the generalization to the Symanzik gauge action, for which
$\mathcal{D}$ itself has non-vanishing off-diagonal elements,
does not introduce any fundamentally new types of contribution to
$\mathcal{T}$. Of course, the expression for $\mathcal{T}$ is much more
involved, but this does not present problems since the expression
is evaluated numerically when doing the loop integral.
This situation is in contrast to what happens if
the action and operators are composed of
thin links, for then $\mathcal{T}$ is diagonal, which greatly
simplifies the resulting expressions.

In order to simplify the expressions for Feynman diagrams,
Ref.~\cite{Lee:2003sk} used the following properties of
$\mathcal{T}_{\mu\nu}$: it is symmetric, and
its off-diagonal elements are
proportional to $\bar s_\mu \bar s_\nu$ multiplied by
a function even in each of the components of $k_\mu$.
For the Wilson gauge action, with diagonal $\mathcal{D}$,
these properties follow from the fact that $h_{\mu\nu}$ has
the same properties.
For the Symanzik gauge action, it turns out that $\mathcal{D}$
also has these properties, from which it is simple to show that
$\mathcal{T}$ does too.
Thus the simplifications used in 
Ref.~\cite{Lee:2003sk} apply for both gauge actions.

We now describe how the analytic formulae of 
Ref.~\cite{Lee:2003sk} must be changed when using
the Symanzik action.\footnote{%
As discussed in Ref.~\cite{Kim:2010fj}, the simple recipe
described here does not work if one uses the asqtad action
because not every diagram can be expressed in terms of the 
composite gluon propagator (due to the presence of the Naik term). 
Some diagrams would need to be calculated anew.}
Two independent approaches to the matching calculation were used
in that work. In the first, explicit expressions were given
for all diagrams (Appendices A, B and C of Ref.~\cite{Lee:2003sk}). 
To obtain the expressions for the Symanzik gauge action one
must make the following replacement:
\begin{equation} 
 \sum_\lambda h_{\mu\lambda}h_{\nu\lambda} \to
 (4 \sum_\rho \bar{s}^2_\rho) \sum_{\alpha\beta} h_{\mu\alpha}h_{\nu\beta} 
 {\cal D}^\text{Imp}_{\alpha\beta}\,.
\end{equation}
In the second method (Appendix D of Ref.~\cite{Lee:2003sk}), 
maximal use was made of the
matching calculation for bilinear operators. For this
part of the calculation, one can simply use our results
for matching factors of bilinear operators with the Symanzik
gauge action~\cite{Kim:2010fj}.
For two classes of diagrams [those of Figs.~\ref{fig:ff2c}(f) and (h)],
bilinear results are not sufficient, and for these
Ref.~\cite{Lee:2003sk} gives explicit expressions.
These are written in terms of the diagonal and
off-diagonal parts of $\mathcal{T}$, and in particular 
in terms of $P_\mu$ and $O_{\mu\nu}$ defined through
\begin{equation}
\mathcal{T}_{\mu\nu} =
\frac{\delta_{\mu\nu} P_\mu
+ (1-\delta_{\mu\nu}) 4 \bar{s}_\mu\bar{s}_\nu O_{\mu\nu}}
{4 \sum_\mu \bar{s}_\mu^2}
\,,
\label{eq:PandO}
\end{equation}
(where repeated indices are not summed).
Here the recipe is to construct $\mathcal{T}_{\mu\nu}$, 
Eq.~(\ref{eq:comp_glu_prop}), using the Symanzik gluon propagator,
use this in Eq.~(\ref{eq:PandO})
to obtain new expressions for $P_\mu$ and $O_{\mu\nu}$, 
and use the latter in the results of Ref.~\cite{Lee:2003sk}.

As in Ref.~\cite{Lee:2003sk}, we evaluate matching
factors using both methods described above and find agreement. This is a
non-trivial check on the numerical implementation of the 
analytic expressions. We have also checked the relations which follow
from Fierz identities and from the $U(1)_\epsilon$ symmetry
of staggered fermions.

\section{Matching factors\label{sec:matching}}


We calculate the matching factors in the usual way by
evaluating the $qq\bar q\bar q$ matrix elements of the
operators both on the lattice and in the continuum, and projecting
onto the different color and spin-taste contributions.
We do so at one-loop order, requiring the evaluation of
the diagrams of
Figs.~\ref{fig:ff1c} and \ref{fig:ff2c} on the lattice.
On the continuum side, only diagrams of types (a) and (g)
contribute, since the continuum four-fermion operators do not
contain gauge fields.
In the continuum calculation one must choose 
the continuation to $4+\epsilon$ dimensions of the operators and
the projectors onto different spin structures.
We follow the conventions described in Refs.~\cite{Lee:2003sk}
and \cite{Gupta:1996yt}.

The matching formula between continuum and lattice-regularized operators
then takes the general form
\begin{equation}
 \mathcal{O}_i^\text{Cont}(\mu)
 = \sum_j {Z}_{ij}(\mu, a)\,
{\mathcal{O}}_j^\text{Lat}(1/a) \,,
\end{equation}
with $\mu$ the continuum regularization scale, and
the lattice spacing now made explicit.
At one-loop order, and with a suitable choice
of lattice operators, the matching factor has the form
\begin{equation}
 {Z}_{ij} = \delta_{ij} +
 \frac{g^2}{(4\pi)^2}
 \bigg[
 - {\gamma}_{ij} \log (\mu a) + {c}_{ij}
 \bigg] 
 + \mathcal{O}(a)\,.
\end{equation}
where ${\gamma}_{ij}$ and ${c}_{ij}$ are, respectively,
the one-loop anomalous dimension matrix and the finite coefficients.
The latter are given by the difference of finite terms in the
continuum and lattice one-loop calculations,
\begin{equation}
 {c}_{ij} 
 = {C}_{ij}^\text{Cont}
 - {C}_{ij}^\text{Lat} \,.
\end{equation}
The general expressions for ${\gamma}_{ij}$
and ${C}_{ij}^\text{Cont}$ are given in Ref.~\cite{Lee:2003sk}
and we do not reproduce them here.\footnote{%
In Ref.~\cite{Lee:2003sk}, a more elaborate notation
is used in which $\gamma$ and $C$ become matrices
in ``color-trace'' space. We do not use this notation here.
We take this opportunity to correct two typographical errors
in Table XIV of Ref.~\cite{Lee:2003sk}: the entries in the
$\hat{\gamma}_{ij}$ column which are $-4$ and $4/3$ 
should be changed to $+4$ and $-4/3$, respectively.}
We only note that
the mixing structure in the continuum is much simpler than
that on the lattice because taste is conserved.

Mean-field improvement of the action and operators
leads to a change in ${C}_{ij}^\text{Lat}$ and
thus in the finite part of the matching factors:
\begin{align}
 {c}_{ij} \stackrel{\rm MF}{\longrightarrow}
 &\ {c}_{ij} - C_F I_{MF} {T}_{ij} \,,
\end{align}
where $C_F = 4/3$, and 
\begin{equation}
 I_{MF} = (4\pi)^2 
 \int^{\pi}_{-\pi}\frac{d^4k}{(2\pi)^4}
 \Big(
 (\bar{s}_2)^2 \mathcal{T}_{11}
 - \bar{s}_1 \bar{s}_2 \mathcal{T}_{12}
 \Big)\,,
\end{equation}
with $\mathcal{T}_{\mu\nu}$ the
composite gluon propagator defined in Eq.~(\ref{eq:comp_glu_prop}).
General results for the mean-field-improvement coefficients,
${T}_{ij}$, can be found in Ref.~\cite{Lee:2001hc},
and are quoted below for the operators considered here.

For each of the indices $i$ and $j$, there are $16^4$ 
choices of the $S$, $F$, $S'$, $F'$. Although lattice symmetries
reduce the number of independent entries, ${c}_{ij}$ remains
a large matrix. We have obtained expressions for all its entries,
but present here only the most interesting subset.

\subsection{Matching Factors for $B_K$\label{subsec:matching:BK}}

The continuum $\Delta S = 2$ four-fermion operator whose matrix element
enters into the kaon mixing parameter $B_K$ is
\begin{equation}
 \mathcal{O}^\text{Cont}_{B_K} =
 [\bar{s}^a \gamma_\mu  (1-\gamma_5) d^a] 
 [\bar{s}^b \gamma_\mu  (1-\gamma_5) d^b] \,.
\end{equation}
In order to calculate 
$\langle \bar K_0|\mathcal{O}^\text{Cont}_{B_K} |K_0\rangle$
using staggered fermions, one must first relate it,
in the continuum, to a matrix element in
an augmented theory in which there are four tastes
for each continuum flavor. In fact, as explained in
Ref.~\cite{Kilcup:1997ye}, 
one also needs to choose the quarks in each
bilinear to have different flavors
[as has been done in the lattice operators defined
in Eqs.~(\ref{eq:1c-tr-op}) and (\ref{eq:2c-tr-op})]. 
This is necessary so that
the Wick contractions in the original and augmented continuum
theories agree. Thus one ends up with an eightfold increase in the
number of valence flavors. To maintain the equivalence of the
sea-quark sectors (and thus the dynamics) of these two theories one
must take the 8th root of the fermion determinant.
This rooting is not controversial in the formal continuum limit, 
and the equality of the corresponding matrix elements in
the two continuum theories is valid non-perturbatively.
We do note, however, that the need for rooting in the
augmented theory implies that this theory is
partially quenched, as has been stressed in Ref.~\cite{Sharpe:2006re}.

The end result, in the ``two spin-trace''
formulation of Refs.~\cite{Sharpe:1986xu,Kilcup:1997ye}, 
is that the relevant operator in the augmented continuum 
theory is (keeping only the positive parity part)
\begin{equation}
 \mathcal{O}_{B_K}^{{\rm Cont}'}
 = \mathcal{O}^{{\rm Cont}'}_{V1}
 + \mathcal{O}^{{\rm Cont}'}_{V2}
 + \mathcal{O}^{{\rm Cont}'}_{A1}
 + \mathcal{O}^{{\rm Cont}'}_{A2}\,,
\label{eq:OBKCont'}
\end{equation}
where
\begin{align}
 \mathcal{O}^{{\rm Cont}'}_{V1} &\equiv 
[\bar S_a (\gamma_\mu \otimes \xi_5) D_b]
[\bar S'_b(\gamma_\mu \otimes \xi_5) D'_a] \,, 
\label{eq:V1Contdef}
\\
 \mathcal{O}^{{\rm Cont}'}_{V2} &\equiv 
[\bar S_a (\gamma_\mu \otimes \xi_5) D_a]
[\bar S'_b(\gamma_\mu \otimes \xi_5) D'_b] \,, 
\label{eq:V2Contdef}
\\
 \mathcal{O}^{{\rm Cont}'}_{A1} &\equiv 
[\bar S_a (\gamma_\mu\gamma_5 \otimes \xi_5) D_b]
[\bar S'_b(\gamma_\mu\gamma_5 \otimes \xi_5) D'_a] \,, 
\label{eq:A1Contdef}
\\
 \mathcal{O}^{{\rm Cont}'}_{A2} &\equiv 
[\bar S_a (\gamma_\mu\gamma_5 \otimes \xi_5) D_a]
[\bar S'_b(\gamma_\mu\gamma_5 \otimes \xi_5) D'_b] \,. 
\label{eq:A2Contdef}
\end{align}
Here $S$, $D$, $S'$ and $D'$ are Dirac fields
having an implicit taste index running over four values.
This index is contracted with the taste matrix $\xi_5$.
The overall normalization of this operator is unimportant
as it cancels in the ratio which defines $B_K$.
Note also that the choice of taste matrix is arbitrary in
the continuum theory---we use $\xi_5$ since that is what is used in 
lattice calculations.

The augmented continuum theory has been chosen to be the
continuum limit of the lattice staggered theory.\footnote{%
Here we assume that rooting introduces no problems with the continuum
limit, following the discussion in
Refs.~\cite{Bernard:2006zw,Sharpe:2006re,Shamir:2006nj,Bernard:2007ma}.}
In particular, at tree level, the operators 
$\mathcal{O}^{{\rm Cont}'}_j$ match onto lattice
operators [defined in Eqs.~(\ref{eq:1c-tr-op}) and (\ref{eq:2c-tr-op})]
as follows:
\begin{align}
 \mathcal{O}^{{\rm Cont}'}_{V1} &\stackrel{{\rm tree}}{=} 
 \mathcal{O}^\text{Lat}_{V1} =
[V_\mu \times P][V_\mu\times P]_I\,,
\\
 \mathcal{O}^{{\rm Cont}'}_{V2} &\stackrel{{\rm tree}}{=} 
 \mathcal{O}^\text{Lat}_{V2} =
[V_\mu \times P][V_\mu\times P]_{II}\,,
\\
 \mathcal{O}^{{\rm Cont}'}_{A1} &\stackrel{{\rm tree}}{=} 
 \mathcal{O}^\text{Lat}_{A1} =
[A_\mu \times P][A_\mu\times P]_I\,,
\\
 \mathcal{O}^{{\rm Cont}'}_{A2} &\stackrel{{\rm tree}}{=} 
 \mathcal{O}^\text{Lat}_{A2} =
[A_\mu \times P][A_\mu\times P]_{II}\,.
\end{align}
Thus the tree-level matching relation for the $B_K$ operator is
\begin{equation}
 \mathcal{O}_{B_K}^{{\rm Cont}'}
 = \mathcal{O}^\text{Lat}_{V1}
  +\mathcal{O}^\text{Lat}_{V2}
  +\mathcal{O}^\text{Lat}_{A1}
  +\mathcal{O}^\text{Lat}_{A2} + O(g^2) + O(a^2)\,.
\end{equation}

At one-loop order, many lattice operators contribute to the matching
formula. It is convenient to divide them into the two classes:
(A) the four operators which arise at tree-level, which have the $\xi_5$
taste matrices in both bilinears, and (B)
the remaining operators, which all turn out to have taste matrices
other than $\xi_5$ in the bilinears.
This division is useful for two reasons.
First, in present numerical calculations only operators
from class (A) are kept, so these are the matching coefficients
that are needed.\footnote{%
The rationale for this is that we use external kaons with taste
$\xi_5$. As shown in Ref.~\cite{VandeWater:2005uq}, however, leaving
out the operators with other tastes leads to an error of ${\cal
O}(\alpha_s m_K^2/\Lambda_\chi^2)$, which is of next-to-leading order
in staggered chiral perturbation theory. This error must be accounted
for when fitting.}
Second, these are the only operators for which
anomalous-dimension matrix elements and finite
continuum coefficients are non-zero.
Thus we write the one-loop matching formula as
\begin{eqnarray}
 \mathcal{O}_{B_K}^{{\rm Cont}'}
 &=& \sum_{i\in (A)} z_i  \mathcal{O}^\text{Lat}_i 
- \frac{g^2}{(4\pi)^2} \sum_{j\in (B)} d^\text{Lat}_j
 \mathcal{O}^\text{Lat}_j \,,
\label{eq:mf_BK}
\\
z_i\! &=& \!
 1 + \frac{g^2}{(4\pi)^2}
 \bigg(\! - 4 \log (\mu a)
  -\frac{11}{3} - d^\text{Lat}_i\bigg)
\label{eq:zi}
\end{eqnarray}
where the subscripts to the sums indicate
that $i$ runs over the four operators in class (A)
while $j$ runs over all operators in class (B).
We have put in the values of the anomalous dimensions
and finite coefficients.
The constants $d^\text{Lat}_j$ are obtained by summing
elements 
of the matrices ${C}^\text{Lat}_{ij}$ introduced above.

\begin{table*}[htbp!] 
\caption[]{Results for $d^\text{Lat}_i$ for various choices of gauge
and fermion action and of the four-fermion operators.}
\label{tab:ff_cLat}
\begin{ruledtabular}
\begin{tabular}{lrrrrrr}
	& $(a)$  &$(b)$  &$(c)$  &$(d)$ & $(e)$ & $(f)$  \\ 
\hline 
Gluon action    & Wilson & Sym & Wilson & Sym & Wilson & Sym \\
Quark action    & Naive & Naive & HYP & HYP & HYP & HYP \\
Mean-field imp. & Y & Y & N & N & Y & Y \\
\hline 
$d^\text{Lat}_{V1}   $   & -2.349(1) & -2.487(1) & -4.984(1) 
                       &-3.649(1) &-2.174(1) &-1.722(1) \\
$d^\text{Lat}_{V2}   $   &-12.915(2) &-11.537(2) &-11.108(2) 
                       &-8.584(2) &-5.487(2) &-4.729(2) \\
$d^\text{Lat}_{A1}   $   & -2.951(1) & -3.077(1) & -5.496(1) 
                       &-4.119(1) &-2.686(1) &-2.192(1) \\
$d^\text{Lat}_{A2}   $   & -3.725(1) & -2.895(1) &  1.012(1) 
                       & 1.087(1) & 1.012(1) & 1.087(1) \\
\hline                                          
Range                   & 10.57 &  9.04 & 12.18 
                       & 9.67 & 6.50 & 5.82 \\  
\end{tabular}
\end{ruledtabular}
\end{table*}

Numerical values for the
$d^\text{Lat}_{1-4}$ are given in table~\ref{tab:ff_cLat}.
We compare the naive staggered action (with operators having
thin links) to the HYP-smeared staggered action (with operators
having smeared links). In the former case, we implement
mean-field improvement (since, in general, perturbation theory
is very poorly convergent without this improvement), while
for the HYP-smeared case we show results both with and without
mean-field improvement. 
For each choice of fermion, we compare the results obtained using
the Wilson and tree-level Symanzik gauge actions, with
the former being from Ref.~\cite{Lee:2003sk}.
We see that improving the gauge action has a small effect,
which, in most cases, reduces the size of the coefficients.
A much more significant reduction is obtained by HYP smearing,
as can be seen by comparing, for example, columns (b) and (f).
(This is the appropriate comparison because both columns show
results with mean-field improvement implemented.)
The results needed for our companion numerical calculation~\cite{Bae:2010ki}
are those of column (f).

As can be seen from Eq.~(\ref{eq:zi}), the full
one-loop correction includes the anomalous dimension and
finite continuum contributions as well as $d^\text{Lat}_i$.
Thus the total size of the one loop correction depends on the
renormalization scale and lattice spacing through the combination
$\mu a$. A better measure of the size of the
correction is the range of the coefficients $d^\text{Lat}_{1-4}$,
from which anomalous dimension and continuum contributions cancel.
The ranges are given in table~\ref{tab:ff_cLat},
and show a small reduction with the use of the improved gauge action.

To give a sense of the numerical size of the matching
coefficients themselves, we show
in table~\ref{tab:mf_BK} results for the $z_{1-4}$
for the ``ultrafine'' MILC lattices ($a\approx0.045\;$fm), 
setting $\mu=1/a$ (``horizontal matching''), 
and using $\alpha_s = g^2/(4\pi) = 0.2096$ 
(the value in the $\overline{\rm MS}$ scheme at $\mu=1/a$).  
For the actions we use in practice [column (f)] 
the one-loop corrections range between $+2\%$ and $-8\%$.

\begin{table}[ht]
\caption[]{Values of $z_i$ for the MILC ultrafine ensembles.
 The notation for actions is as in table~\protect\ref{tab:ff_cLat}.}
 \label{tab:mf_BK}
\begin{ruledtabular}
\begin{tabular}{lrrrrrr}
	& $(a)$  &$(b)$  &$(c)$  &$(d)$ & $(e)$ & $(f)$ \\ 
\hline 
$z_{V1}$   & 0.978 & 0.980  & 1.022 & 1.000 & 0.975 &  0.968 \\
$z_{V2}$   & 1.154 & 1.131  & 1.124 & 1.082 & 1.030 &  1.018 \\
$z_{A1}$   & 0.988 & 0.990  & 1.031 & 1.008 & 0.984 &  0.975 \\
$z_{A2}$   & 1.001 & 0.987  & 0.922 & 0.921 & 0.922 &  0.921 \\
\end{tabular}
\end{ruledtabular}
\end{table}

To give a complete view of the one-loop matching, we present,
in Tables~\ref{tab:BKrest} and \ref{tab:BKrest-2}, 
the coefficients $d^\text{Lat}_j$
for all other operators which appear at this order.
We see that improving the gauge action leads, as above, to
a small reduction in the magnitude
of all the matching coefficients.
Note that, since these mixings are pure lattice artifacts,
with no anomalous dimension or other continuum contributions,
reducing the size of the coefficients is an unambiguous 
improvement.

As noted above, in present numerical calculations the mixing
with these operators is not being incorporated in the lattice
operators, but rather is
a source of systematic error that must be estimated by fitting.
An alternative approach would be to include the
dominant operators from the Tables in the one-loop matched operator.
As one can see, there are relatively few operators having
$O(1)$ coefficients:
\begin{enumerate}
\item $[S\times V_\mu][S\times V_\mu]$,
\item $[P\times V_\mu][P\times V_\mu]$ and $[P\times V_\mu][P\times V_\nu]$,
\item $[T_{\mu\nu}\times V_\mu][T_{\mu\nu}\times V_\mu]$,
$[T_{\mu\nu}\times V_\rho][T_{\mu\nu}\times V_\rho]$\\ 
$[T_{\mu\nu}\times V_\mu][T_{\mu\nu}\times V_\nu]$ and
$[T_{\mu\nu}\times V_\rho][T_{\mu\nu}\times V_\eta]$.
\end{enumerate}
The remainder of the coefficients are an order of magnitude or
more smaller (i.e. $|d^\text{Lat}_i|<0.2$). 
Since these coefficients are multiplied by
$g^2/(4\pi)^2 \approx 0.017-0.025$ for $a\approx 0.045-0.12\;$fm,
we expect the contributions to $B_K$ from the remaining operators to
be very small.

\begin{table}[ht]
\caption{Matching coefficients $d^\text{Lat}_j$
[defined in Eq.~(\ref{eq:mf_BK})]
for the operator required for calculating $B_K$,
and for operators $j$ which have different
taste than the continuum operator (\ref{eq:OBKCont'}).
Lattice operators and fermion action are HYP-smeared and the gauge action
is either Wilson---column (c)---or Symanzik---column (d).
Results in column (c) are obtained from
Tables I-IV of Ref.~\cite{Lee:2003sk}.
The coefficients $T_{j}$ give the impact of mean-field improvement: 
$d^\text{Lat}_j \to d^\text{Lat}_j + C_F I_{MF} T_j$,
with $I_{MF}=1.053786$ for the Wilson gauge action
and $I_{MF}=0.722795$ for the Symanzik gauge action.
Greek indices are implicitly summed, 
with the condition that they are unequal,
and the further constraint that for the
operator $[V_\mu \times T_{\nu\rho}][V_\mu\times T_{\nu\rho}]$,
$\nu<\rho$, while for the operators
$[T_{\mu\nu}\times V_\rho][T_{\mu\nu}\times V_\rho]$ and
$[T_{\mu\nu}\times V_\rho][T_{\mu\nu}\times V_\eta]$,
$\mu<\nu$.
Results are accurate to at least $\pm 2$ in the last digit quoted.}
\label{tab:BKrest}
\begin{ruledtabular}
\begin{tabular}{ l c c c c }
${\cal O}^{Lat}_j $ & color trace & (c) & (d) & $T_j$ \\
\hline							     
$[ S \times V_\mu ][ S \times V_\mu ] $ & I
& $-3.450$ & $-2.805$ & $1$ \\
$[ S \times V_\mu ][ S \times V_\mu ] $ & II
& $-0.263$ & $-0.249$ & $0$ \\
$[ S \times V_\mu ][ S \times V_\nu ] $ & I
& $0.118$ & $0.108$ & $0$ \\
$[ S \times V_\mu ][ S \times V_\nu ] $ & II
& $-0.104$ & $-0.097$ & $0$ \\
$[ S \times A_\mu ][ S \times A_\mu ] $ & I
& $0.043$ & $0.028$ & $0$ \\
$[ S \times A_\mu ][ S \times A_\mu ] $ & II
& $-0.052$ & $-0.035$ & $0$ \\
$[ S \times A_\mu ][ S \times A_\nu ] $ & I
& $-0.015$ & $-0.010$ & $0$ \\
$[ S \times A_\mu ][ S \times A_\nu ] $ & II
& $0.002$ & $0.002$ & $0$ \\
$[ V_\mu \times S ][ V_\mu \times S ] $ & I
& $-0.044$ & $-0.029$ & $0$ \\
$[ V_\mu \times S ][ V_\mu \times S ] $ & II
& $-0.008$ & $-0.005$ & $0$ \\
$[ V_\mu \times T_{\mu\nu} ][ V_\mu \times T_{\mu\nu} ] $ & I
& $-0.124$ & $-0.086$ & $0$ \\
$[ V_\mu \times T_{\mu\nu} ][ V_\mu \times T_{\mu\nu} ] $ & II
& $-0.114$ & $-0.084$ & $0$ \\
$[ V_\mu \times T_{\mu\nu} ][ V_\mu \times T_{\mu\rho} ] $ & I
& $0.029$ & $0.023$ & $0$ \\
$[ V_\mu \times T_{\mu\nu} ][ V_\mu \times T_{\mu\rho} ] $ & II
& $-0.023$ & $-0.019$ & $0$ \\
$[ V_\mu \times T_{\mu\nu} ][ V_\mu \times T_{\nu\rho} ] $ & I
& $0.016$ & $0.014$ & $0$ \\
$[ V_\mu \times T_{\mu\nu} ][ V_\mu \times T_{\nu\rho} ] $ & II
& $0.002$ & $0.002$ & $0$ \\
$[ V_\mu \times T_{\nu\rho} ][ V_\mu \times T_{\mu\nu} ] $ & I
& $0.016$ & $0.014$ & $0$ \\
$[ V_\mu \times T_{\nu\rho} ][ V_\mu \times T_{\mu\nu} ] $ & II
& $0.002$ & $0.002$ & $0$ \\
$[ V_\mu \times T_{\nu\rho} ][ V_\mu \times T_{\nu\rho} ] $ & I
& $-0.118$ & $-0.091$ & $0$ \\
$[ V_\mu \times T_{\nu\rho} ][ V_\mu \times T_{\nu\rho} ] $ & II
& $-0.037$ & $-0.030$ & $0$ \\
$[ V_\mu \times T_{\nu\rho} ][ V_\mu \times T_{\nu\eta} ] $ & I
& $0.027$ & $0.022$ & $0$ \\
$[ V_\mu \times T_{\nu\rho} ][ V_\mu \times T_{\nu\eta} ] $ & II
& $-0.020$ & $-0.016$ & $0$ \\
$[ T_{\mu\nu} \times V_\mu ][ T_{\mu\nu} \times V_\mu ] $ & I
& $2.071$ & $1.547$ & $-1$ \\
$[ T_{\mu\nu} \times V_\mu ][ T_{\mu\nu} \times V_\mu ] $ & II
& $-0.538$ & $-0.485$ & $0$ \\
$[ T_{\mu\nu} \times V_\mu ][ T_{\mu\nu} \times V_\nu ] $ & I
& $-0.410$ & $-0.383$ & $0$ \\
$[ T_{\mu\nu} \times V_\mu ][ T_{\mu\nu} \times V_\nu ] $ & II
& $0.452$ & $0.417$ & $0$ \\
$[ T_{\mu\nu} \times V_\mu ][ T_{\mu\nu} \times V_\rho ] $ & I
& $0.129$ & $0.120$ & $0$ \\
$[ T_{\mu\nu} \times V_\mu ][ T_{\mu\nu} \times V_\rho ] $ & II
& $0.126$ & $0.118$ & $0$ \\
$[ T_{\mu\nu} \times V_\rho ][ T_{\mu\nu} \times V_\mu ] $ & I
& $0.129$ & $0.120$ & $0$ \\
$[ T_{\mu\nu} \times V_\rho ][ T_{\mu\nu} \times V_\mu ] $ & II
& $0.126$ & $0.118$ & $0$ \\
$[ T_{\mu\nu} \times V_\rho ][ T_{\mu\nu} \times V_\rho ] $ & I
& $-2.930$ & $-2.331$ & $1$ \\
$[ T_{\mu\nu} \times V_\rho ][ T_{\mu\nu} \times V_\rho ] $ & II
& $-0.346$ & $-0.316$ & $0$ \\
$[ T_{\mu\nu} \times V_\rho ][ T_{\mu\nu} \times V_\eta ] $ & I
& $0.652$ & $0.610$ & $0$ \\
$[ T_{\mu\nu} \times V_\rho ][ T_{\mu\nu} \times V_\eta ] $ & II
& $-0.153$ & $-0.143$ & $0$ \\
\end{tabular}
\end{ruledtabular}
\end{table}
\begin{table}[ht]
\caption{Matching coefficients $d^\text{Lat}_j$
(continued from Table~\ref{tab:BKrest}).
Again, Greek indices are implicitly summed, 
with the condition that they are unequal,
with the further constraint that for the
operator $[A_\mu \times T_{\nu\rho}][A_\mu\times T_{\nu\rho}]$,
$\nu<\rho$, and for the operators
$[T_{\mu\nu}\times A_\rho][T_{\mu\nu}\times A_\rho]$ and
$[T_{\mu\nu}\times A_\rho][T_{\mu\nu}\times A_\eta]$,
$\mu<\nu$.
}

\label{tab:BKrest-2}
\begin{ruledtabular}
\begin{tabular}{ l c c c c }
${\cal O}^{Lat}_j $ & color trace & (c) & (d) & $T_j$ \\
\hline
$[ T_{\mu\nu} \times A_\mu ][ T_{\mu\nu} \times A_\mu ] $ & I
& $-0.026$ & $-0.017$ & $0$ \\
$[ T_{\mu\nu} \times A_\mu ][ T_{\mu\nu} \times A_\mu ] $ & II
& $-0.045$ & $-0.030$ & $0$ \\
$[ T_{\mu\nu} \times A_\mu ][ T_{\mu\nu} \times A_\nu ] $ & I
& $-0.010$ & $-0.007$ & $0$ \\
$[ T_{\mu\nu} \times A_\mu ][ T_{\mu\nu} \times A_\nu ] $ & II
& $-0.003$ & $-0.002$ & $0$ \\
$[ T_{\mu\nu} \times A_\mu ][ T_{\mu\nu} \times A_\rho ] $ & I
& $0.001$ & $0.000$ & $0$ \\
$[ T_{\mu\nu} \times A_\mu ][ T_{\mu\nu} \times A_\rho ] $ & II
& $-0.002$ & $-0.001$ & $0$ \\
$[ T_{\mu\nu} \times A_\rho ][ T_{\mu\nu} \times A_\mu ] $ & I
& $0.001$ & $0.000$ & $0$ \\
$[ T_{\mu\nu} \times A_\rho ][ T_{\mu\nu} \times A_\mu ] $ & II
& $-0.002$ & $-0.001$ & $0$ \\
$[ T_{\mu\nu} \times A_\rho ][ T_{\mu\nu} \times A_\rho ] $ & I
& $-0.068$ & $-0.046$ & $0$ \\
$[ T_{\mu\nu} \times A_\rho ][ T_{\mu\nu} \times A_\rho ] $ & II
& $-0.024$ & $-0.016$ & $0$ \\
$[ T_{\mu\nu} \times A_\rho ][ T_{\mu\nu} \times A_\eta ] $ & I
& $0.000$ & $-0.000$ & $0$ \\
$[ T_{\mu\nu} \times A_\rho ][ T_{\mu\nu} \times A_\eta ] $ & II
& $0.003$ & $0.002$ & $0$ \\
$[ A_\mu \times S ][ A_\mu \times S ] $ & I
& $-0.003$ & $-0.002$ & $0$ \\
$[ A_\mu \times S ][ A_\mu \times S ] $ & II
& $-0.022$ & $-0.014$ & $0$ \\
$[ A_\mu \times T_{\mu\nu} ][ A_\mu \times T_{\mu\nu} ] $ & I
& $-0.124$ & $-0.086$ & $0$ \\
$[ A_\mu \times T_{\mu\nu} ][ A_\mu \times T_{\mu\nu} ] $ & II
& $-0.114$ & $-0.084$ & $0$ \\
$[ A_\mu \times T_{\mu\nu} ][ A_\mu \times T_{\mu\rho} ] $ & I
&  $0.022$ & $0.017$ & $0$ \\
$[ A_\mu \times T_{\mu\nu} ][ A_\mu \times T_{\mu\rho} ] $ & II
& $-0.002$ & $-0.002$ & $0$ \\
$[ A_\mu \times T_{\mu\nu} ][ A_\mu \times T_{\nu\rho} ] $ & I
& $-0.004$ & $-0.003$ & $0$ \\
$[ A_\mu \times T_{\mu\nu} ][ A_\mu \times T_{\nu\rho} ] $ & II
& $0.011$ & $0.009$ & $0$ \\
$[ A_\mu \times T_{\nu\rho} ][ A_\mu \times T_{\mu\nu} ] $ & I
& $-0.004$ & $-0.003$ & $0$ \\
$[ A_\mu \times T_{\nu\rho} ][ A_\mu \times T_{\mu\nu} ] $ & II
& $0.011$ & $0.009$ & $0$ \\
$[ A_\mu \times T_{\nu\rho} ][ A_\mu \times T_{\nu\rho} ] $ & I
& $-0.106$ & $-0.083$ & $0$ \\
$[ A_\mu \times T_{\nu\rho} ][ A_\mu \times T_{\nu\rho} ] $ & II
& $-0.074$ & $-0.055$ & $0$ \\
$[ A_\mu \times T_{\nu\rho} ][ A_\mu \times T_{\nu\eta} ] $ & I
& $-0.017$ & $-0.015$ & $0$ \\
$[ A_\mu \times T_{\nu\rho} ][ A_\mu \times T_{\nu\eta} ] $ & II
& $0.011$ & $0.009$ & $0$ \\
$[ P \times V_\mu ][ P \times V_\mu ] $ & I
& $2.566$ & $2.004$ & $-1$ \\
$[ P \times V_\mu ][ P \times V_\mu ] $ & II
& $-0.547$ & $-0.503$ & $0$ \\
$[ P \times V_\mu ][ P \times V_\nu ] $ & I
& $0.151$ & $0.141$ & $0$ \\
$[ P \times V_\mu ][ P \times V_\nu ] $ & II
& $0.326$ & $0.307$ & $0$ \\
$[ P \times A_\mu ][ P \times A_\mu ] $ & I
& $-0.063$ & $-0.041$ & $0$ \\
$[ P \times A_\mu ][ P \times A_\mu ] $ & II
& $-0.042$ & $-0.028$ & $0$ \\
$[ P \times A_\mu ][ P \times A_\nu ] $ & I
& $0.012$ & $0.008$ & $0$ \\
$[ P \times A_\mu ][ P \times A_\nu ] $ & II
& $-0.005$ & $-0.003$ & $0$ \\
\end{tabular}
\end{ruledtabular}
\end{table}

\subsection{Matching Factors for other four-fermion operators\label{subsec:matching:other}}

Models of new physics can lead, after integrating out heavy particles,
to $\Delta S=2$ operators with different Dirac structure from that
in $\mathcal{O}^\text{Cont}_{B_K}$.
To constrain these models one needs to know the matrix elements of these new
operators. A standard basis is~\cite{Gabbiani:1996hi}
\begin{eqnarray}
 \mathcal{O}^\text{Cont}_{2} &=&
 [\bar{s}^a (1-\gamma_5) d^a] [\bar{s}^b (1-\gamma_5) d^b] \,,
\\
 \mathcal{O}^\text{Cont}_{3} &=&
 [\bar{s}^a (1-\gamma_5) d^b] [\bar{s}^b (1-\gamma_5) d^a] \,,
\\
 \mathcal{O}^\text{Cont}_{4} &=&
 [\bar{s}^a (1-\gamma_5) d^a] [\bar{s}^b (1+\gamma_5) d^b] \,,
\\
 \mathcal{O}^\text{Cont}_{5} &=&
 [\bar{s}^a (1-\gamma_5) d^b] [\bar{s}^b (1+\gamma_5) d^a] \,.
\end{eqnarray}
In this section we present one-loop matching coefficients for
these operators.\footnote{%
Linear combinations of these operators are also needed for
calculating the $K\to\pi$ matrix elements of the $I=3/2$ part
of the electromagnetic penguin contribution to $\epsilon'/\epsilon$.
In the context of staggered fermions this is explained in 
Ref.~\cite{Kilcup:1997ye}.}

As for $\mathcal{O}^\text{Cont}_{B_K}$, the first step is to
match the operators into the augmented continuum theory.
The result is
\begin{align}
 \mathcal{O}_{2}^{{\rm Cont}'} &=
 \mathcal{O}^{{\rm Cont}'}_{S2}+
 \mathcal{O}^{{\rm Cont}'}_{P2}+
\nonumber \\ &\ \
 - \frac12 \left(\mathcal{O}^{{\rm Cont}'}_{S1}+
 \mathcal{O}^{{\rm Cont}'}_{P1}-
 \mathcal{O}^{{\rm Cont}'}_{T1}\right)\,,
\\
 \mathcal{O}_{3}^{{\rm Cont}'} &=
 \mathcal{O}^{{\rm Cont}'}_{S1}+
 \mathcal{O}^{{\rm Cont}'}_{P1}+
\nonumber \\ &\ \
 - \frac12 \left(\mathcal{O}^{{\rm Cont}'}_{S2}+
 \mathcal{O}^{{\rm Cont}'}_{P2}-
 \mathcal{O}^{{\rm Cont}'}_{T2}\right)\,,
\\
 \mathcal{O}_{4}^{{\rm Cont}'} &=
 \mathcal{O}^{{\rm Cont}'}_{S2}-
 \mathcal{O}^{{\rm Cont}'}_{P2}+
\nonumber \\ &\ \
 - \frac12 \left(\mathcal{O}^{{\rm Cont}'}_{V1}-
 \mathcal{O}^{{\rm Cont}'}_{A1}\right)\,,
\\
 \mathcal{O}_{5}^{{\rm Cont}'} &=
 \mathcal{O}^{{\rm Cont}'}_{S1}-
 \mathcal{O}^{{\rm Cont}'}_{P1}+
\nonumber \\ &\ \
 - \frac12 \left(\mathcal{O}^{{\rm Cont}'}_{V2}-
 \mathcal{O}^{{\rm Cont}'}_{A2}\right)\,,
\end{align}
where some of the operators are defined in 
Eqs.~(\ref{eq:V1Contdef}-\ref{eq:A2Contdef}), and the others are
\begin{align}
 \mathcal{O}^{{\rm Cont}'}_{S1} &\equiv 
[\bar S_a ({\bf 1} \otimes \xi_5) D_b]
[\bar S'_b({\bf 1} \otimes \xi_5) D'_a] \,, 
\\
 \mathcal{O}^{{\rm Cont}'}_{S2} &\equiv 
[\bar S_a ({\bf 1} \otimes \xi_5) D_a]
[\bar S'_b({\bf 1} \otimes \xi_5) D'_b] \,, 
\\
 \mathcal{O}^{{\rm Cont}'}_{P1} &\equiv 
[\bar S_a (\gamma_5 \otimes \xi_5) D_b]
[\bar S'_b(\gamma_5 \otimes \xi_5) D'_a] \,, 
\\
 \mathcal{O}^{{\rm Cont}'}_{P2} &\equiv 
[\bar S_a (\gamma_5 \otimes \xi_5) D_a]
[\bar S'_b(\gamma_5 \otimes \xi_5) D'_b] \,, 
\\
 \mathcal{O}^{{\rm Cont}'}_{T1} &\equiv 
\sum_{\mu<\nu} [\bar S_a (\gamma_\mu\gamma_\nu \otimes \xi_5) D_b]
[\bar S'_b(\gamma_\mu\gamma_\nu \otimes \xi_5) D'_a] \,, 
\\
 \mathcal{O}^{{\rm Cont}'}_{T2} &\equiv 
\sum_{\mu<\nu} [\bar S_a (\gamma_\mu\gamma_\nu \otimes \xi_5) D_a]
[\bar S'_b(\gamma_\mu\gamma_\nu \otimes \xi_5) D'_b] \,.
\end{align}
At tree-level, the new operators in the augmented
continuum theory match onto lattice operators in the obvious way:
\begin{align}
 \mathcal{O}^{{\rm Cont}'}_{S1} &\stackrel{{\rm tree}}{=} 
 \mathcal{O}^\text{Lat}_{S1} =
[S \times P][S \times P]_I\,,
\\
 \mathcal{O}^{{\rm Cont}'}_{S2} &\stackrel{{\rm tree}}{=} 
 \mathcal{O}^\text{Lat}_{S2} =
[S \times P][S \times P]_{II}\,,
\\
 \mathcal{O}^{{\rm Cont}'}_{P1} &\stackrel{{\rm tree}}{=} 
 \mathcal{O}^\text{Lat}_{P1} =
[P \times P][P \times P]_I\,,
\\
 \mathcal{O}^{{\rm Cont}'}_{P2} &\stackrel{{\rm tree}}{=} 
 \mathcal{O}^\text{Lat}_{P2} =
[P \times P][P \times P]_{II}\,,
\\
 \mathcal{O}^{{\rm Cont}'}_{T1} &\stackrel{{\rm tree}}{=} 
 \mathcal{O}^\text{Lat}_{T1} =
\sum_{\mu<\nu}[T_{\mu\nu} \times P][T_{\mu\nu}\times P]_I\,,
\\
 \mathcal{O}^{{\rm Cont}'}_{T2} &\stackrel{{\rm tree}}{=} 
 \mathcal{O}^\text{Lat}_{T2} =
\sum_{\mu<\nu}[T_{\mu\nu} \times P][T_{\mu\nu}\times P]_{II}\,.
\end{align}

As for $\mathcal{O}_{B_K}^{{\rm Cont}'}$,
each of the operators $\mathcal{O}_{2-5}^{{\rm Cont}'}$
matches at one loop order onto class (A) lattice operators
composed of bilinears with taste $\xi_5$, and class (B) operators having other
tastes. Continuum mixing involves only class (A) operators---mixing
with operators of class (B) is a lattice effect.
We display here only results for mixing with class (A) operators,
for several reasons. First, these are likely to be
the subset of operators used in numerical simulations. Second,
the contribution of class (B) operators is of 
next-to-next-to-leading order in staggered chiral perturbation
theory, using the power counting of Ref.~\cite{VandeWater:2005uq}. 
This is because the contribution is suppressed both by $\alpha_s$,
and by the need to have a chiral loop due to the mismatch
between the taste of the operator and external states. Unlike for
$\mathcal{O}_{B_K}^{{\rm Cont}'}$, there is no chiral enhancement
to raise the contribution to next-to-leading order.
Finally, we do not show results for class (B) mixing for the sake of
brevity.

We write the one-loop matching formula as 
\begin{equation}
 \mathcal{O}_{i}^{{\rm Cont}'}
 = \sum_{j\in (A)} z_{ij}  \mathcal{O}^\text{Lat}_{j} 
- \frac{g^2}{(4\pi)^2} \sum_{k\in (B)} d^\text{Lat}_{ik}
 \mathcal{O}^\text{Lat}_k \,,
\label{eq:mf_Oj}
\end{equation}
with $i=2-4$ and
\begin{eqnarray}
z_{ij} &=& b_{ij} + \frac{g^2}{(4\pi)^2}
 \Big( - \gamma_{ij} \log (\mu a) + d^\text{Cont}_{ij}
  - d^\text{Lat}_{ij} 
  \nonumber \\
& & \hspace{20mm} - C_F I_{MF} T_{ij}\Big)\,.
\label{eq:zij}
\end{eqnarray}
We stress that we are using a different basis for the
continuum and lattice operators, so that the tree-level
contribution is no longer $\delta_{ij}$. Because of this,
we denote the finite parts as $d_{ij}$ rather than $C_{ij}$.
The $d^\text{Cont}_{ij}$ matrix comes from Ref.~\cite{Lee:2003sk}.
The elements of $d^\text{Lat}_{ij}$, which are calculated
numerically, are new results.
The $T_{ij}$ term is present only if the
operator is mean-field improved. Numerical values for
$I_{MF}$ are given in the caption of Table~\ref{tab:BKrest}.

Results for the four operators are presented in
Tables~\ref{tab:matchO2}-\ref{tab:matchO5}.
We show results for HYP-smeared operators without
mean-field improvement, but
include the values of $T_{ij}$
so that mean-field improvement can be easily implemented.

\begin{table}[htbp!] 
\caption[]{Matching coefficients 
entering Eq.~(\ref{eq:zij}) for $i=2$,
i.e. for $\mathcal{O}^{{\rm Cont}'}_2$. 
The finite lattice coefficients are for the HYP-smeared fermion action
and operators, and either (c) the Wilson gauge action or (d) the
Symanzik gauge action (following the labeling used in
Table~\ref{tab:ff_cLat}. 
The $d^\text{Cont}_{ij}$ matrix comes from Ref.~\cite{Lee:2003sk}.
Results are accurate to the number of
digits quoted.}
\label{tab:matchO2}
\begin{ruledtabular}
\begin{tabular}{lrrrrrr}
Operator $j$ & $b_{2j}$  &$\gamma_{2j}$  &$d^\text{Cont}_{2j}$  
&$d^\text{Lat}_{2j} (c)$ & $d^\text{Lat}_{2j} (d)$ & $T_{2j}$ \\
\hline 
$\mathcal{O}^\text{Lat}_{S1}$   
& $-1/2$ & $6$   & $-59/12$ &   2.70 &   2.34 & -1 \\
$\mathcal{O}^\text{Lat}_{S2}$   
& $1$    & $-10$ & $+73/12$ & -17.80 & -14.53 &  6 \\
$\mathcal{O}^\text{Lat}_{P1}$   
& $-1/2$ & $6$   & $-59/12$ &   3.64 &   3.17 & -1 \\
$\mathcal{O}^\text{Lat}_{P2}$   
& $1$    & $-10$ & $+73/12$ &   5.42 &   4.06 & -2 \\
$\mathcal{O}^\text{Lat}_{T1}$   
& $1/2$& $-14/3$ & $+29/12$ &  -2.94 &  -2.52 &  1 \\
$\mathcal{O}^\text{Lat}_{T2}$
& $0$    & $2/3$ &  $-5/4$  &   0.03 &   0.01 &  0 \\
\end{tabular}
\end{ruledtabular}
\end{table}

\begin{table}[htbp!] 
\caption[]{Matching coefficients 
entering Eq.~(\ref{eq:zij}) for $i=3$.
Notation as in Table~\ref{tab:matchO2}.}
\label{tab:matchO3}
\begin{ruledtabular}
\begin{tabular}{lrrrrrr}
Operator $j$ & $b_{3j}$  &$\gamma_{3j}$  &$d^\text{Cont}_{3j}$  
&$d^\text{Lat}_{3j} (c)$ & $d^\text{Lat}_{3j} (d)$ & $T_{3j}$ \\
\hline 
$\mathcal{O}^\text{Lat}_{S1}$   
& $1$    & $8$   &  $1/3$ &  -4.35 &  -2.88 &  2 \\
$\mathcal{O}^\text{Lat}_{S2}$   
& $-1/2$ & $0$   & $-5/3$ &   7.13 &   5.38 & -3 \\
$\mathcal{O}^\text{Lat}_{P1}$   
& $1$    & $8$   &  $1/3$ &  -6.23 &  -4.56 &  2 \\
$\mathcal{O}^\text{Lat}_{P2}$   
& $-1/2$ & $0$   & $-5/3$ &  -0.25 &  -0.14 &  1 \\
$\mathcal{O}^\text{Lat}_{T1}$   
& $0$    & $8/3$ & $-1$   &   0.14 &   0.24 &  0 \\
$\mathcal{O}^\text{Lat}_{T2}$
& $1/2$  &$16/3$ & $-1/3$ &  -2.31 &  -1.55 &  1 \\
\end{tabular}
\end{ruledtabular}
\end{table}

\begin{table}[htbp!] 
\caption[]{Matching coefficients 
entering Eq.~(\ref{eq:zij}) for $i=4$.
Notation as in Table~\ref{tab:matchO2}.}
\label{tab:matchO4}
\begin{ruledtabular}
\begin{tabular}{lrrrrrr}
Operator $j$ & $b_{4j}$  &$\gamma_{4j}$  &$d^\text{Cont}_{4j}$  
&$d^\text{Lat}_{4j} (c)$ & $d^\text{Lat}_{4j} (d)$ & $T_{4j}$ \\
\hline 
$\mathcal{O}^\text{Lat}_{S1}$   
& $0$    & $0$   &  $-3$ &   0    &   0    &  0 \\
$\mathcal{O}^\text{Lat}_{S2}$   
& $1$    & $-16$ & $+23/3$ & -19.12 & -16.02 &  6 \\
$\mathcal{O}^\text{Lat}_{P1}$   
& $0$    & $0$   &  $+3$ &   0    &   0    &  0 \\
$\mathcal{O}^\text{Lat}_{P2}$   
& $-1$   & $16$  & $-23/3$ &  -6.93 &  -5.09 &  2 \\
$\mathcal{O}^\text{Lat}_{V1}$   
& $-1/2$ & $8$   & $-23/6$   &   3.02 &   2.72 & -1 \\
$\mathcal{O}^\text{Lat}_{V2}$
& $0$    & $0$   & $+3/2$ &   0.37 &   0.34 &  0 \\
$\mathcal{O}^\text{Lat}_{A1}$   
& $1/2$  & $-8$  & $+23/6$   &  -3.27 &  -2.95 &  1 \\
$\mathcal{O}^\text{Lat}_{A2}$
& $0$    & $0$   & $-3/2$ &   0.40 &   0.37 &  0 \\
\end{tabular}
\end{ruledtabular}
\end{table}

\begin{table}[htbp!] 
\caption[]{Matching coefficients 
entering Eq.~(\ref{eq:zij}) for $i=5$.
Notation as in Table~\ref{tab:matchO2}.}
\label{tab:matchO5}
\begin{ruledtabular}
\begin{tabular}{lrrrrrr}
Operator $j$ & $b_{5j}$  &$\gamma_{5j}$  &$d^\text{Cont}_{5j}$  
&$d^\text{Lat}_{5j} (c)$ & $d^\text{Lat}_{5j} (d)$ & $T_{5j}$ \\
\hline 
$\mathcal{O}^\text{Lat}_{S1}$   
& $1$    & $2$   &  $1/6$ &  -4.67 &  -3.42 &  2 \\
$\mathcal{O}^\text{Lat}_{S2}$   
& $0$    & $-6$  & $-1/2$ &  -2.32 &  -2.45 &  0 \\
$\mathcal{O}^\text{Lat}_{P1}$   
& $-1$   & $-2$  & $-1/6$ &   6.55 &   5.10 & -2 \\
$\mathcal{O}^\text{Lat}_{P2}$   
& $0$    & $6$   & $ 1/2$ &  -3.32 &  -2.59 &  0 \\
$\mathcal{O}^\text{Lat}_{V1}$   
& $0$    & $3$   & $ 1/4$ &   0.16 &   0.27 &  0 \\
$\mathcal{O}^\text{Lat}_{V2}$
& $-1/2$ & $-1$  & $-1/12$&   5.24 &   4.04 & -2 \\
$\mathcal{O}^\text{Lat}_{A1}$   
& $0$    & $-3$  & $-1/4$ &  -0.16 &  -0.27 &  0 \\
$\mathcal{O}^\text{Lat}_{A2}$
& $1/2$  & $1$   & $ 1/12$&   0.05 &   0.09 &  0 \\
\end{tabular}
\end{ruledtabular}
\end{table}

The tables show that, for coefficients with magnitudes larger than
about 5, improvement of the gauge action leads to a small reduction
in the size of the finite lattice coefficients.\footnote{%
The range of the coefficients is not useful for these
operators as the contributions from anomalous dimensions differ.}
Even with this improvement,
we see that the largest lattice coefficients have a magnitude as
large as 16, although most have a magnitude smaller than 10.
These numbers are larger than those we found
for the $B_K$ operator (Table~\ref{tab:ff_cLat}).
On a coarse lattice with $a\approx 0.12\;$fm,
and $g^2/(4\pi)^2\approx 0.025$, the largest one-loop term could
give a 40\% correction. This is not a precise statement because
we have not included the contribution from finite continuum terms
and from the anomalous dimensions.
Nevertheless, it is a warning to expect perturbation theory to
be less convergent for the operators $\mathcal{O}_{2-5}$ than
for $\mathcal{O}_{B_K}$.

The situation is better if one implements mean-field improvement.
As can be seen from the Tables, this reduces all the larger coefficients,
so that the largest magnitude is now $\approx 10$.
Since mean-field improvement is relatively straightforward to implement,
our results suggest that this would be worth the required investment.

Finally, we note that, unlike for $\mathcal{O}_{B_K}$, one finds
very large finite coefficients if one uses mean-field improved naive
fermions (with either gauge action). The largest magnitudes are $\approx 50$,
indicating a complete breakdown in the convergence of perturbation theory.
Because of this, we do not include the results in the tables.

\section{Conclusion\label{sec:conclude}}

We have calculated the matching factors for four-fermion operators
using various fermion and gauge actions. Most useful are our results
for the fermion action and operators constructed using HYP-smeared
links with the Symanzik improved gluon action. These are needed for
our ongoing calculation of $B_K$ and related matrix 
elements~\cite{Bae:2010ki,Yoon:2010bm}. 
For these operators, the one-loop corrections
are of moderate size for the $B_K$ operator,
with the range of corrections being $\approx 10
\times \alpha_s/(4\pi) \approx \alpha_s$, which is the naively
expected size. The same holds true for the operators induced by new physics,
as long as one implements mean-field improvement.
For all operators, we find that
improving the gauge action generically leads to a reduction
in the size of one-loop matching
coefficients, but that the effect is relatively small.

\section{Acknowledgments}
The research of W.~Lee is supported by the Creative Research
Initiatives program (3348-20090015) of the NRF grant funded by the
Korean government (MEST).
The work of S.~Sharpe is supported in part by the US DOE grant
no.~DE-FG02-96ER40956.
%
\bibliographystyle{apsrev} 
\bibliography{ref} 

\end{document}